\documentclass[aps,prd,reprint,superscriptaddress]{revtex4-1}
\usepackage{amsmath}
\usepackage{amssymb}
\usepackage{graphicx}
\usepackage[caption=false]{subfig}
\usepackage{enumitem}
\usepackage{color}
\usepackage[percent]{overpic}
\usepackage{bm}
\usepackage{rotating}
\usepackage{hyperref}
\usepackage{cancel}
\usepackage{comment}
\usepackage{braket}
\usepackage[normalem]{ulem}

\newcommand{\abs}[1]{\left| #1 \right|} 

\newcommand{\ii}{\mathrm{i}}

\newcommand{\dd}{\mathrm{d}}

\begin{document}

\title{
$\hat{\bm{p}}\cdot\hat{\bm{A}}$ vs $\hat{\bm{x}}\cdot\hat{\bm{E}}$: Gauge invariance in quantum optics}
\author{Nicholas Funai}
\affiliation{Institute for Quantum Computing, University of Waterloo, Waterloo, Ontario, N2L 3G1, Canada}
\affiliation{Department of Applied Mathematics, University of Waterloo, Waterloo, Ontario, N2L 3G1, Canada}
\author{Jorma Louko}
\affiliation{School of Mathematical Sciences, University of Nottingham, Nottingham NG7 2RD, United Kingdom}
\author{Eduardo Mart\'{i}n-Mart\'{i}nez}
\affiliation{Institute for Quantum Computing, University of Waterloo, Waterloo, Ontario, N2L 3G1, Canada}
\affiliation{Department of Applied Mathematics, University of Waterloo, Waterloo, Ontario, N2L 3G1, Canada}
\affiliation{Perimeter Institute for Theoretical Physics, 31 Caroline St N, Waterloo, Ontario, N2L 2Y5, Canada}


\begin{abstract}
We compare the predictions of the fundamentally motivated minimal coupling ($\hat{\bm{p}}\cdot\hat{\bm{A}}$) and the ubiquitous dipole coupling ($\hat{\bm{x}}\cdot\hat{\bm{E}}$) in the light-matter interaction. By studying the light-matter interaction for hydrogen-like atoms we find that the dipole approximation cannot be a-priori justified to analyze the physics of vacuum excitations (a very important phenomenon in relativistic quantum information) since a dominant wavelength is absent in those problems, no matter how small (as compared to any frequency scale) the atom is. Remarkably, we show that the dipole approximation in those regimes can still be valid as long as the interaction time is longer than the light-crossing time of the atoms, which is a very reasonable assumption. We also highlight some of the subtleties that one has to be careful with when working with the explicitly gauge noninvariant  nature of the minimal coupling and we compare it with the explicitly gauge invariant dipole coupling.


\end{abstract}

\maketitle

\section{Introduction}

Whilst quantum theory was still in its infancy there was a desire to integrate Schr\"{o}dinger's equation into the might of electromagnetism. This was not only motivated by the want of a theoretical framework for the quantum dynamics of charged particles, but also by the need to design improved experiments to test atom-field interactions, some of which ultimately lead to the laser \cite{MAIMAN1960}. One of the main challenges in this formalism was how to deal with the gauge freedom of the electromagnetic (EM) field and ensure that physically measurable quantities, such as transition rates, also respected this freedom. The constraining effect of the gauge freedom when combined with a local gauge freedom for the electron wavefunction lead to the derivation of the minimally coupled Hamiltonian \cite{Scully}
\begin{align}
\begin{split}
\ii\frac{\partial}{\partial t}\psi(\bm{x},t)&=\left\{
\vphantom{\frac{1}{2\mu_{e}}\left(\hat{\bm{p}}-q\bm{A}(\bm{x},t)\right)^{2}+V(\bm{x})+qU(\bm{x},t)}
\frac{1}{2\mu_{e}}\left(\hat{\bm{p}}-q\bm{A}(\bm{x},t)\right)^{2}\right.\\
+&\left.\vphantom{\frac{1}{2\mu_{e}}\left(\hat{\bm{p}}-q\bm{A}(\bm{x},t)\right)^{2}+V(\bm{x})+qU(\bm{x},t)}V(\bm{x})+qU(\bm{x},t)\right\}\psi(\bm{x},t),
\end{split}\label{eq1}
\end{align}
where the wavefunctionn local gauge transformation is given by $\tilde{\psi}(\bm{x},t)=e^{-\ii q \chi(\bm{x},t)}\psi(\bm{x},t)$ so that \eqref{eq1} is invariant under the usual EM vector and scalar potentials gauge transformations
\begin{align}
\tilde{\bm{A}}(\bm{x},t)&=\bm{A}(\bm{x},t)-\nabla\chi(\bm{x},t),\label{g1}\\
\tilde{U}(\bm{x},t)&=U(\bm{x},t)+\dot{\chi}(\bm{x},t).\label{g2}
\end{align}
Here $\mu_{e}$ refers to the reduced mass of the electron-nucleus system.

This atom-field interaction seems to resolve the gauge issue introduced by the EM field; however, historically the improper use of different gauges predicted different transition rates; which, if correct, would have allowed an experimentalist to isolate a physical gauge. In order to resolve this gauge issue in 1931 G\"{o}ppert-Mayer \cite{1931AnP...401..273G} wrote down the electric dipole coupling, which approximates \eqref{eq1} by
\begin{align}
\ii\frac{\partial}{\partial t}\psi(\bm{x},t)=\left\{\frac{1}{2\mu_{e}}\hat{\bm{p}}^{2}+V(\bm{x})+q\hat{\bm{x}}\cdot\bm{E}(\bm{x},t)\right\}\psi(\bm{x},t),\label{eq4}
\end{align}
and makes Schr\"{o}dinger's equation free from any gauge terms. The validity of this coupling required the dipole approximation \cite{BECKER1985107}, an approximation that requests that all plane-waves in the EM field with wavevector $\bm{k}$ must obey $R\abs{\bm{k}}\ll1 $ where $R$ is the characteristic length of the support of the electron wavefunction (e.g. Bohr radius). Under these circumstances the EM field is approximately spatially constant over the size of the atom. In addition to its simplicity, this interaction was consolidated over time by experiments \cite{PhysRev.85.259} and is confidently used today \cite{PhysRevA.35.4253,DELLANNO200653}.

The electric dipole coupling introduced by G\"{o}ppert-Mayer has worked well for the experimental regimes accessible in the 20th century; however, the purpose built minimally coupled Hamiltonian defying its gauge independence by producing gauge dependent observable predictions was still a problem that required a solution. The solution to this problem was produced by Lamb, Schlicher and Scully \cite{PhysRevA.36.2763} with the introduction of `physical observables'. The issue of gauge noninvariance originated from a misuse of gauge transformations via a change of EM gauge without performing the corresponding wavefunction gauge transformation. In order to prevent such ambiguity all `physical observables' would be defined as functions of the position operator $\hat{\bm{x}}$ and the `mechanical momentum', i.e. a form of gauge covariant derivative given by $\hat{\bm{\pi}}=\hat{\bm{p}}-q \bm{A}(\bm{x},t)$. Using the mechanical momentum, Schr\"{o}dinger's equation \eqref{eq1} becomes
\begin{align}
\ii\frac{\partial}{\partial t}\psi(\bm{x},t)&=\left\{\frac{1}{2\mu_{e}}\hat{\bm{\pi}}^{2}(\bm{x},t)+V(\bm{x})+qU(\bm{x},t)\right\}\psi(\bm{x},t).\label{eq5.1}
\end{align}

In particular this concept of `mechanical' or `physical' observables also restricts the set of measurable attributes; for example the expectation value of momentum $\braket{\hat{\bm{p}}}=\braket{\psi|\hat{\bm{p}}|\psi}$ is no longer an allowed measurement as a gauge transformation of the state $\ket{\psi}$ reveals a gauge dependence on the expectation value. Conversely the expectation value of `physical' observables, e.g. $\braket{\hat{\bm{\pi}}}$, are allowed measurements as any gauge transformation of the EM field produce terms that are exactly balanced by the associated wavefunction gauge transformations. This enforces that any `physical' observable produced measurements are gauge invariant, a necessary condition for any prediction that has to match experimental results. 

In summary, gauge invariant transition amplitudes must be formulated as those between the eigenstates of the free `mechanical' Hamiltonian,
\begin{align}
\hat{H}=\frac{1}{2\mu_{e}}\hat{\bm{\pi}}^{2}(\bm{x},t)+V(\bm{x}) . \label{mechH}
\end{align}
As $\hat{\bm{\pi}}=\hat{\bm{p}}-q \bm{A}(\bm{x},t)$ and $\hat{\bm{p}}=-i\nabla$, these eigenstates depend on~$\bm{A}$, 
and they will in general not coincide with the eigenstates of the $q=0$ Hamiltonian. 
The point however is that the eigenstates of $\hat{H}$ can be expanded perturbatively in $q$ about the eigenstates of the $q=0$ Hamiltonian: 
this choice for the eigenstates of $\hat{H}$ provides a perturbative expansion of gauge invariant transition amplitudes that have a physical meaning, 
as the observable transition amplitudes between eigenstates of the $q=0$ Hamiltonian.
In particular Lamb et al. \cite{PhysRevA.36.2763} demonstrated that by following this approach and by using point-like atoms in classical EM fields (where the electric dipole approximation holds) the minimally coupled Hamiltonian yielded the same transition probabilities as the electric dipole Hamiltonian. However; to prove this, it was assumed in \cite{PhysRevA.36.2763} that the atoms are point-like, again relying on an approximation that requires a \textit{dominant wavelength} (or, rather, a range of dominant/relevant wavelengths) of the field to be much larger than the atomic size  e.g., a coherent excitation of the field of peak wavelength much larger than the size of the atom, or a process of spontaneous emission where the gap of the atom has an associated wavelength again much larger than the atomic radius.

 In this paper we will study to what extent the dipole approximation and the minimal coupling (with the correct gauge considerations) coincide for finite-size hydrogen-like atoms when no notion of \textit{dominant wavelength} (or range of dominant wavelengths) is present, e.g. vacuum excitations: ground state atoms in the presence of the vacuum of the quantum fields interacting for finite times. We will be following a similar approach to Lamb et al. in order to properly compare the electric dipole coupling with the minimally coupled Hamiltonian with particular attention to why the electric dipole coupling is so frequently consistent with experimental results.

 We wish to clarify that we will consider the dipole expansion only. Higher order terms can be added to improve the multipole approximation as shown by \cite{PhysRevA.28.2649}; however we will compare the minimal model with the dipole model, the simplest approximation and the most commonly used; and show under what conditions the dipole model can still be used even in the absence of dominant wavelengths.
 
 Our motivation is to validate the use of the dipole approximation for relativistic quantum information scenarios; in particular where EM fields are treated quantumly and where rotating wave-type approximations are no longer appropriate, denying us a means of defining a \textit{dominant wavelength}, i.e. beyond the assumptions of previous works. This is the case when we study phenomena where the initial state of the system is the field vacuum and the ground state of the atom (such as e.g., the Fermi problem \cite{RevModPhys.4.87,PhysRevLett.107.150402} or vacuum entanglement harvesting \cite{VALENTINI1991321,Reznik,Pozas2015}, above all when they are computed modelling realistic hydrogen-like atoms \cite{PhysRevD.94.064074}). 
 
 To summarize, the work presented here extends the domain of the work by Lamb et al.\ by removing the assumption of a dominant wavelength, allowing us to treat a wider range of physical situations. Our work uses finite sized atoms, quantum fields and the full Hydrogen atom Hilbert space; none of which were used by Lamb et al. In our analysis we do not consider dissipative losses, only concentrating on Schr\"{o}dinger's equation and the effects of the different coupling type. In the discussion we also raise the issue of switching the interaction on and off and its effects of the dipole approximation.

\section{The models}

When treating the electric dipole coupling \eqref{eq4} we note that the Hamiltonian is explicitly gauge invariant. The full dipole model then consists of leaving the electron wavefunction invariant under EM gauge transformations in order to guarantee gauge invariant dynamics. This is a common approach adopted by the quantum optics community, particularly exploiting this electron gauge invariance to define their [preparation and measurement] basis of states as eigenstates of $\frac{1}{2\mu_{e}}\hat{\bm{p}}^{2}+V(\hat{\bm{x}})$ \cite{ABATE1974269}. 

Conversely, as stated above, the minimal coupling Hamiltonian \eqref{eq1} is not gauge invariant; therefore, we adopt a local gauge transformation of the electron wavefunction to guarantee gauge invariant physical quantities. These local gauge transformations introduce the migraine of deciding how to define  the preparation and measurement basis of states. Any choice of basis will be mathematically correct, provided the relevant gauge transformations are properly applied; however, experimental constraints would ultimately arbitrate. Lamb et al.'s suggestion was to use eigenstates of the mechanical energy, which dynamically change with the field state; although, due to energy level degeneracies these states do not seem to define an appropriate preparation and measurement basis consistent with the usual $\ket{n,l,m}$ atomic states commonly used in light-matter interactions. This issue is not addressed by Lamb et al. as they restricted their atoms to 2-level systems.

Note our units are $\hbar=c=\epsilon_{0}=1$. On some occasions these constants will be explicitly restored for clarity.

\subsection{The states}
In following the spirit of Lamb et al. \cite{PhysRevA.36.2763} we work perturbatively to first order in $q$ and define a `dressed' state (assuming, for now, a classical field theory)
\begin{align}
\ket{\tilde{\psi}_{t,l}}=\sum_{k}\left(\delta_{lk}+\ii q L_{lk}(t)\right)\ket{\psi_{k}},\label{eq5}
\end{align}
where (in our case) $\ket{\psi_{k}}$ are $\ket{n,l,m}$ eigenstates of the standard atomic Hamiltonian $\frac{1}{2\mu_{e}}\hat{\bm{p}}^{2}+V(r)$, orbital angular momentum $\hat{L}^{2}$ and z-orbital angular momentum $\hat{L}_{z}$ respectively. The coefficients $L_{lk}$ will be functions of the EM potentials and are chosen such that the dressed state $\ket{\tilde{\psi}_{t,l}}$ is `gauge' invariant, i.e. applying the wavefunction local gauge transformation also gauge transforms the EM potentials defining $L_{lk}$, all to first order in a $q$ perturbative expansion. 

Since $\ket{\tilde{\psi}_{t,l}}\rightarrow \ket{\psi_{l}}$ as $q\rightarrow 0$ and $\ket{\psi_{l}}$ is an eigenstate of the free Hamiltonian $ {\hat{\bm{p}}^{2}}/{2\mu_{e}+V} $ then in order to attain gauge invariance one could set $\ket{\tilde{\psi}_{t,l}}$ to be an eigenstate of $ {\hat{\bm{\pi}}^{2}}/{2\mu_{e}}+V $, as suggested by Lamb et al. If all the eigenenergies of the free Hamiltonian are non-degenerate, this is accomplished by 
\begin{align}
L_{lk}(t)=-\frac{\braket{\psi_{k}|\frac{\bm{A}(\hat{\bm{x}},t)\cdot\hat{\bm{p}}+\hat{\bm{p}}\cdot\bm{A}(\hat{\bm{x}},t)}{2\mu_{e}}|\psi_{l}}}{\ii(E_{l}-E_{k})}. 
\end{align}
If some eigenenergies are degenerate this idea has, however, a technical difficulty because first order perturbation theory muddles the energy eigenstates of the free Hamiltonian within the degenerate sectors already at zeroth order, and we would lose our notion of `perturbing' the free states $\ket{n,l,m}$ defined by energy and angular momentum.
A convenient way to avoid this issue is to choose
\begin{align}
L_{lk}(t)=\begin{cases}
-\frac{\braket{\psi_{k}|\frac{\bm{A}(\hat{\bm{x}},t)\cdot\hat{\bm{p}}+\hat{\bm{p}}\cdot\bm{A}(\hat{\bm{x}},t)}{2\mu_{e}}|\psi_{l}}}{\ii(E_{l}-E_{k})} & \text{if $E_{l}\neq E_{k}$},\\
-\bra{\psi_{k}}\int\limits_{0}^{t}\dd s\, U(\hat{\bm{x}},s)\ket{\psi_{l}} &\text{if $E_{l}=E_{k}$}, 
\end{cases}\label{eq6}
\end{align}
which ensures gauge invariance of the transition probabilities under the minimal coupling.  Stronger than that, in fact this already  ensures  the gauge invariance of the transition amplitudes, as we show in Appendix~\ref{seca05}. Note that as $q\rightarrow 0$ these states return to the usual atomic $\ket{n,l,m}$ eigenstates. 

One can interpret $L_{lk}(t)$ as defining the boundary conditions for the perturbative solution of Schr\"{o}dinger's equation, i.e. initial state $\ket{\tilde{\psi}_{0,i}}$ and final measurement state $\ket{\tilde{\psi}_{T,f}}$, measured at time $T$. Physically, these dressed states form an orthonormal basis whose measurements produce physical quantities in the sense that the transition rates are independent of the choice of gauge (shown in appendix \ref{seca05}).

\subsection{The equations}

The soon to be quantised Schr\"{o}dinger equation in the Coulomb gauge,
\begin{align}
\ii\frac{\partial\psi(\bm{x},t)}{\partial t}&=\left\{\frac{1}{2\mu_{e}}\left(\hat{\bm{p}}-q\bm{A}(\bm{x},t)\right)^{2}+V(r)\right\}\psi(\bm{x},t),
\end{align}
to which we apply the gauge transformation (c.f.\ Scully and Zubairy \cite{Scully} and also appendix \ref{seca0})
\begin{align}
\tilde{\psi}(\bm{x},t)&=e^{-\ii q\left(\bm{A}(\bm{x},t)\cdot\bm{x}\right)}\psi(\bm{x},t)
\end{align}
finally yields (See appendix \ref{seca0})
\begin{align}
\begin{split}
\ii\frac{\partial\tilde{\psi}(\bm{x},t)}{\partial t}
&=\left\{\frac{1}{2\mu_{e}}\left(\hat{\bm{p}}+q\left[\left(x_{i}\nabla\right)A_{i}(\bm{x},t)\right]\right)^{2}
\vphantom{\frac{1}{2\mu_{e}}\left(\hat{\bm{p}}+q\left[\left(x_{i}\nabla\right)A_{i}(\bm{x},t)\right]\right)^{2}-q\hat{\bm{x}}\cdot\bm{E}(\bm{x},t)+V(r)}\right.\\
-&\left.\vphantom{\frac{1}{2\mu_{e}}\left(\hat{\bm{p}}+q\left[\left(x_{i}\nabla\right)A_{i}(\bm{x},t)\right]\right)^{2}-q\bm{x}\cdot\bm{E}(\bm{x},t)+V(r)}q\bm{x}\cdot\bm{E}(\bm{x},t)+V(r)\right\}\tilde{\psi}(\bm{x},t).
\end{split}\label{eq9}
\end{align}
Notice the term $(x_{i}\nabla)A_{i}(\bm{x},t)$; this term can be considered small if the variations of $A_{i}$ are small on the length scale of the support of the wavefunction, in Fourier terms this means dominant Fourier modes of $A_{i}$ must obey the relation $a_{0} k\ll 1$ where $a_{0}$ is the length scale of the support of the wavefunction (for Hydrogen orbitals $a_{0}$ is the Bohr radius). This is a slight reformulation of the dipole approximation. We therefore have the two equations of motion for the minimal and dipole models respectively,
\begin{align}
\ii\frac{\partial\psi(\bm{x},t)}{\partial t}&=\left\{\frac{1}{2\mu_{e}}\left(\hat{\bm{p}}-q\bm{A}(\bm{x},t)\right)^{2}+V(r)\right\}\psi(\bm{x},t),\label{eq10}\\
\ii\frac{\partial\tilde{\psi}(\bm{x},t)}{\partial t}
&=\left\{\frac{1}{2\mu_{e}}\hat{\bm{p}}^{2}-q\bm{x}\cdot\bm{E}(\bm{x},t)+V(r)\right\}\tilde{\psi}(\bm{x},t).\label{eq11}
\end{align}

At this point we depart from the semiclassical model and instate $\bm{A}$ and $U$ as members of the fully relativistic 4-potential for EM field with quantum degrees of freedom. As shown in appendix \ref{seca05} the transition rates between dressed states is gauge invariant and so we rewrite \eqref{eq6}, \eqref{eq10} and \eqref{eq11} with a quantised EM field. Specifically, the Coulomb gauge for the EM field obeys
\begin{align}
\bm{\nabla}\cdot\hat{\bm{A}}&=0,\\
\hat{U}&=0,
\end{align}
such that
\begin{align}
\begin{split}
\hat{\bm{A}}(\bm{x},t)&=\int\frac{\dd^{3}\bm{k}}{(2\pi)^{3/2}\sqrt{2\omega}}\sum_{\lambda=1}^{2} \bm{\epsilon}_{\lambda}(\bm{k})\left(\hat{a}_{\lambda}^{\vphantom{\dagger}} (\bm{k})e^{-\ii(\omega t-\bm{k}\cdot\bm{x})}\right.\\
+&\left.\hat{a}_{\lambda}^{\dagger}(\bm{k})e^{\ii(\omega t-\bm{k}\cdot\bm{x})}\right),
\end{split}\\
\bm{k}\cdot\bm{\epsilon}_{\lambda}(\bm{k})&=0,\\
\hat{\bm{E}}(\bm{x},t)&=-\frac{\partial}{\partial t}\hat{\bm{A}}(\bm{x},t),
\end{align}
and $\hat{a}_{\lambda}^{\dagger}(\bm{k}),\hat{a}_{\lambda}^{\vphantom{\dagger}}(\bm{k})$ are the plane-wave mode creation and annihilation operators respectively.

Given that we are working in an interaction picture where the EM free field evolution is encoded into the field operators, we can write the following decomposition of the total Hamiltonian (first order perturbation theory):
\begin{align}
\hat{H}&=\hat{H}_{0}+q\hat{H}_{\text{I}}+O(q^{2}),\label{eq15}\\
\hat{H}_{0}&=\frac{1}{2\mu_{e}}\hat{\bm{p}}^{2}+V(r),\\
\hat{H}_{\text{I}}^{\text{min}}&=-\frac{1}{\mu_{e}}\hat{\bm{A}}(\bm{x},t)\cdot\hat{\bm{p}},\\
\hat{H}_{\text{I}}^{\text{dip}}&=-\hat{x}\cdot\hat{\bm{E}}(\bm{x},t).
\end{align}
For completeness the fully quantum dressed states take the form
\begin{align}
\ket{\tilde{\psi}_{t,l},\phi_{i}}=\sum_{k}\left(\delta_{lk}+\ii q \hat{L}_{lk}(t)\right)\ket{\psi_{k}}\ket{\phi_{i}},\label{eq5.5}
\end{align}
where $\ket{\phi_{i}}$ is the state of the EM field and for the minimal case
\begin{align}
\hat{L}_{lk}(t)=\begin{cases}
-\frac{\braket{\psi_{k}|\frac{\hat{\bm{A}}(\hat{\bm{x}},t)\cdot\hat{\bm{p}}}{\mu_{e}}|\psi_{l}}}{\ii(E_{l}-E_{k})} & \text{if $E_{l}\neq E_{k}$},\\
0 &\text{if $E_{l}=E_{k}$},
\end{cases}\label{eq6.5}
\end{align}
since we are working in the Coulomb gauge and $\hat{L}_{lk}=0$ for the dipole case.

The main concern now is whether the quantum nature of the EM field permits the existence of a `dominant wavelength' and if this wavelength satisfies the dipole approximation. This is particularly relevant when we study phenomena where the initial state of the system is the field vacuum and the ground state of the atom, such as is commonly studied in relativistic quantum information.


\section{Dynamics and transition probabilities}
\subsection{General field state}
Our setup presumes an ability to prepare the electron in a dressed state described by \eqref{eq5.5} and \eqref{eq6.5} at time $t=0$; and to projectively measure the state in a dressed basis also described by \eqref{eq5.5} and \eqref{eq6.5} at some final time $t=T$. Since we are working to first order in perturbation theory, the wavefunction of the electron can be perturbatively represented by
\begin{align}
\ket{\tilde{\psi}_{l}(t),\phi_{i}}&=\sum_{k}\left(\delta_{lk}+\ii q\hat{K}_{lk}(t)\right)e^{-\ii E_{k}t}\ket{\psi_{k}}\ket{\phi_{i}},
\end{align}
i.e. $\ket{\tilde{\psi}_{l}(t),\phi}$ would be a time evolved representation of the dressed state corresponding to the undressed state $\ket{\psi_{l}}\ket{\phi}$. Note that $\hat{K}_{lk}(0)=\hat{L}_{lk}(0)$.

As shown in appendix \ref{seca2}, Schr\"{o}dinger's equation yields
\begin{align}
\dot{\hat{K}}_{lk}(t)&=-\braket{\psi_{k}|\hat{H}_{1}|\psi_{l}}e^{\ii(E_{k}-E_{l})t},
\end{align}
and therefore
\begin{align}
\hat{K}_{lk}(T)&=-\int\limits_{0}^{T}\dd t\braket{\psi_{k}|\hat{H}_{1}|\psi_{l}}e^{\ii(E_{k}-E_{l})t}+\hat{L}_{lk}(0).
\end{align}

These two equations give us a complete description of the wavefunction at time $T$. Now our focus turns to computing the probability amplitude of measuring the state of the system in $\ket{\tilde{\psi}_{T,f},\phi_{f}}$, which becomes
\begin{align}
\begin{split}
&\braket{\tilde{\psi}_{T,f},\phi_{f}|\tilde{\psi}_{i}(T),\phi_{i}}\\
&=\ii q\bra{\phi_{f}}\left(\hat{K}_{if}(T)e^{-\ii E_{f}T}-\hat{L}_{if}(T)e^{-\ii E_{i}T}\right)\ket{\phi_{i}},
\end{split}
\end{align}
where we have used the property $\hat{L}_{lk}^{\dagger}=\hat{L}_{kl}$. Note that we will not measure the field $\ket{\phi_{f}}$, instead we will trace over the field to attain a final probability amplitude.

After some computation these inner products can be compressed into simplified expressions
\begin{widetext}
\begin{align}
\braket{\tilde{\psi}_{T,f},\phi_{f}|\tilde{\psi}_{i}(T),\phi_{i}}_{\text{min}}&=T\frac{q}{\mu_{e}}e^{-\ii E_{f}T}\int\frac{\dd^{3}\bm{k}}{(2\pi)^{3/2}}\sqrt{\frac{\omega}{2}}\sum_{\lambda=1}^{2}\bm{\epsilon}_{\lambda}(k)\cdot\bra{\phi_{f}}\left(\hat{a}_{\lambda}^{\vphantom{\dagger}}e^{\ii(\Omega-\omega)\frac{T}{2}}\text{sinc}\left((\Omega-\omega)\frac{T}{2}\right)\braket{\psi_{f}|\frac{e^{\ii\bm{k}\cdot\bm{x}}\nabla}{\Omega}|\psi_{i}}\right.\nonumber\\
-&\left.
\hat{a}_{\lambda}^{\dagger}e^{\ii(\Omega+\omega)\frac{T}{2}}\text{sinc}\left((\Omega+\omega)\frac{T}{2}\right)\braket{\psi_{f}|\frac{e^{-\ii\bm{k}\cdot\bm{x}}\nabla}{\Omega}|\psi_{i}}\right)\ket{\phi_{i}},
\label{eq25}\\
\braket{\tilde{\psi}_{T,f},\phi_{f}|\tilde{\psi}_{i}(T),\phi_{i}}_{\text{dip}}&=
-Tqe^{-\ii E_{f}T}\int\frac{\dd^{3}\bm{k}}{(2\pi)^{3/2}}\sqrt{\frac{\omega}{2}}\sum_{\lambda=1}^{2}\bm{\epsilon}_{\lambda}(k)\cdot\bra{\phi_{f}}\left(\hat{a}_{\lambda}^{\vphantom{\dagger}}e^{\ii(\Omega-\omega)\frac{T}{2}}\text{sinc}\left((\Omega-\omega)\frac{T}{2}\right)\braket{\psi_{f}|e^{\ii\bm{k}\cdot\bm{x}}\hat{\bm{r}}|\psi_{i}}\right.\nonumber\\
-&\left.\hat{a}_{\lambda}^{\dagger}
e^{\ii(\Omega+\omega)\frac{T}{2}}\text{sinc}\left((\Omega+\omega)\frac{T}{2}\right)\braket{\psi_{f}|e^{-\ii\bm{k}\cdot\bm{x}}\hat{\bm{r}}|\psi_{i}}\right)\ket{\phi_{i}},\label{eq26}
\end{align}
\end{widetext}
where $\text{sinc}(x)=\sin(x)/x$, $\Omega=E_{f}-E_{i}$ and $\mu_{e}$ is the electron-proton reduced mass. Of particular importance here is the sinc term. This introduces a weak polynomial type decay with increasing $\omega$. This decay is a consequence of the `sudden switching' of the interaction between the atom and the EM field. The weakness of this decay is the source of the difference between the dipole and minimal models.

The form of \eqref{eq25} and \eqref{eq26} encourage us to define new variables for derivational simplicity
\begin{align}
&\braket{\tilde{\psi}_{T,f},\phi_{f}|\tilde{\psi}_{i}(T),\phi_{i}}\nonumber\\
=&\bra{\phi_{f}}\int\,\dd^{3}\bm{k}\sum_{\lambda=1}^{2}\left(h_{1,\lambda}\hat{a}_{\lambda}^{\vphantom{\dagger}}+h_{2,\lambda}\hat{a}_{\lambda}^{\dagger}\right)\ket{\phi_{i}},
\end{align}
where $h_{1,\lambda}$ and $h_{2,\lambda}$ are chosen to match up with \eqref{eq25} and \eqref{eq26} for each of the Hamiltonians under investigation. Using this compact expression we can determine the probability of transition
\begin{align}
\begin{split}
&\sum_{\phi_{f}}\abs{\braket{\tilde{\psi}_{T,f},\phi_{f}|\tilde{\psi}_{i}(T),\phi_{i}}}^{2}\\
=&\sum_{\lambda,\lambda'=1}^{2}\bra{\phi_{i}}\int\dd^{3}\bm{k}\left(h_{1,\lambda}\hat{a}_{\lambda}^{\vphantom{\dagger}}+h_{2,\lambda}\hat{a}_{\lambda}^{\dagger}\right)^{\dagger}\\
\times&\sum_{\phi_{f}}\ket{\phi_{f}}\bra{\phi_{f}}\int\dd^{3}\bm{k}'\left(h_{1,\lambda'}\hat{a}_{\lambda'}^{\vphantom{\dagger}}+h_{2,\lambda'}\hat{a}_{\lambda'}^{\dagger}\right)\ket{\phi_{i}}.
\end{split}
\end{align}
Now $\sum_{\phi_{f}}\ket{\phi_{f}}\bra{\phi_{f}}=\mathbb{I}$ is the resolution of the identity for fields, therefore the probability of transition from initial to final states is
\begin{align}
\begin{split}
&P(i\rightarrow f)\\
&=\sum_{\lambda,\lambda'=1}^{2}\left<\int\dd^{3}\bm{k}\int\dd^{3}\bm{k}'\left(h^{*}_{1,\lambda}(\bm{k})\hat{a}_{\lambda}^{\dagger}(\bm{k})+h^{*}_{2,\lambda}(\bm{k})\hat{a}_{\lambda}^{\vphantom{\dagger}}(\bm{k})\right)\right.\\
\times&\left.\left(h_{1,\lambda'}(\bm{k}')\hat{a}_{\lambda'}^{\vphantom{\dagger}}(\bm{k}')+h_{2,\lambda'}(\bm{k}')\hat{a}_{\lambda'}^{\dagger}(\bm{k}')\right)\right>_{\phi_{i}}.
\end{split}
\end{align}

This expression can be further simplified by exploiting the commutation relations of the field operators
\begin{align}
\begin{split}
&P(i\rightarrow f)\\
&\!\!\!\!\!\!=\overbrace{\left<:\sum_{\lambda,\lambda'=1}^{2}\int\dd^{3}\bm{k}\int\dd^{3}\bm{k}'\left(h^{*}_{1,\lambda}(\bm{k})\hat{a}_{\lambda}^{\dagger}(\bm{k})+h^{*}_{2,\lambda}(\bm{k})\hat{a}_{\lambda}^{\vphantom{\dagger}}(\bm{k})\right)\right.}^{P_{\phi}}\\
\times&\left.\left(h_{1,\lambda'}(\bm{k}')\hat{a}_{\lambda'}^{\vphantom{\dagger}}(\bm{k}')+h_{2,\lambda'}(\bm{k}')\hat{a}_{\lambda'}^{\dagger}(\bm{k}')\right):\vphantom{\sum_{\lambda=1}^{2}}\right>_{\phi_{i}}\\
+&\underbrace{\sum_{\lambda=1}^{2}\int\dd^{3}\bm{k}h_{2,\lambda}^{*}(\bm{k})h_{2,\lambda}(\bm{k})}_{P_{0}},
\end{split}\label{eq30}
\end{align}
where the colons indicate normal ordering. In this sense we detach the vacuum contributions ($P_{0}$) from excited field contributions ($P_{\phi}$). A quick inspection of \eqref{eq30} shows that for $\ket{\phi_{i}}=\ket{0}$ the $P_{\phi}$ term will be zero and the $P_{0}$ term remains. Also by inspection we note that the $P_{0}$ term is an integral and sum over non-negative numbers, in contrast to the $P_{\phi}$ term that is the product of sums and integrals over the complex plane; and so we intuite that the $P_{0}$ term will be significant for many cases and the $P_{\phi}$ term becomes relevant in very specific cases or for reasonably strong EM field excitations.

\subsection{Vacuum excitation and spontaneous emission}\label{sec3b}

The natural first step is to compare predictions of the two models for $\ket{\phi_{i}}=\ket{0}$, i.e. the vacuum state. Under these circumstances we attempt to compute \eqref{eq30}. For the transition $1s\rightarrow 2p$ we have (shown in Appendix \ref{seca2})
\begin{align}
\label{eq31} P_{\text{dip}}&=\frac{262144\hbar^{3}\epsilon_{0}}{177147 c^{3}q^{2}Z^{2}\mu_{e}^{2}}\\
\nonumber&\qquad\qquad\times\int_0^\infty\!\!\!\!\dd\omega\frac{\omega^{3}}{(1+\frac{4 a_{0}^{2}}{9c^{2}Z^{2}}\omega^{2})^{6}}\frac{\sin^{2}\left((\omega+\Omega)\frac{t}{2}\right)}{\left((\omega+\Omega)\frac{1}{2}\right)^{2}},\\
P_{\text{min}}&=\frac{262144\hbar^{3}\epsilon_{0}}{177147 c^{3}q^{2}Z^{2}\mu_{e}^{2}} \label{eq32}\\
\nonumber&\qquad\qquad \times\int_0^\infty\!\!\!\!\dd\omega\frac{\omega^{3}}{(1+\frac{4 a_{0}^{2}}{9c^{2}Z^{2}}\omega^{2})^{4}}\frac{\sin^{2}\left((\omega+\Omega)\frac{t}{2}\right)}{\left((\omega+\Omega)\frac{1}{2}\right)^{2}},
\end{align}
where $q$ is the electron charge, $Z$ is the proton number of this Rydberg type atom, $\mu_{e}$ is the reduced mass of the system.  Here we have reintroduced fundamental constants for completeness. We have also used $\Omega=E_{f}-E_{i}$, which for Rydberg atoms is given by 
\begin{equation}
    E_{f}-E_{i}=\frac12\mu_{e} Z^{2} \alpha^{2}\left(\frac{1}{n_{i}^{2}}-\frac{1}{n_{f}^{2}}\right),
\end{equation} and $\alpha=q^{2}/4\pi$ (in natural units); where $\alpha$ is the fine structure constant, $n_{i}$ is the principal quantum number of the initial state and $n_{f}$ is the principal quantum number of the final state. 

\eqref{eq31} and \eqref{eq32} have identical coefficients and one significant difference, namely a different decay rate of the integrand with respect to $\omega$, which generates the discrepency between the two couplings. If high frequency contributions could be dampened then these two integrands could be well approximated by one another. This observation suggests more general conditions for the two models to predict the same probabilities.

Indeed, in this form, it is easy to see why when there's a dominant frequency, the dipole model approximates the minimal one for long times: Consider if $\Omega<0$ (atom initially in the excited state), then for $\omega=-\Omega$ we have resonnance and consequently, for long times, one can apply Fermi's Golden rule (related to the single mode approximation) of the form \begin{align} \lim_{t\rightarrow\infty}\frac{\sin^{2}(\eta t/2)}{(\eta/2)^{2}t}=\pi\delta(\eta/2).\end{align}
In this case $\Omega$ becomes the field's `dominant' frequency and the dipole approximation criterion becomes $\Omega a_{0}/Z\ll 1$. Such a condition, when coupled with the relevant zeroth order Taylor expansion of \eqref{eq31} and \eqref{eq32} yields equal predictions from both couplings. Notice, however, that if $\Omega>0$ then the transition is an excitation and the sinc contribution does not resonate (i.e. $\nexists \omega\geq 0$ such that $\omega+\Omega=0$). In other words, this single-mode like approximation would not be justified if we were looking at the vacuum excitation probability of the field for finite times.

However, this golden rule/single mode approximation is not alone in isolating a single mode or ranges of modes that dominate EM field behaviour. Generally light-matter interactions may include intrinsic field UV cutoffs, time dependence in the interaction strength or secondary non-radiative processes introducing non-perturbative time dependences; any of which, we argue, could, in principle, be used to satisfy the dipole approximation ($\omega a_{0}/Z\ll 1$), and hence the zeroth order Taylor expansions of \eqref{eq31} and \eqref{eq32} become
\begin{align}
P_{\text{dip}}\rightarrow&\frac{262144\hbar^{3}\epsilon_{0}}{177147 c^{3}q^{2}Z^{2}\mu_{e}^{2}}\int\limits_{0}^{\omega\ll \frac{Z}{a_{0}}}\dd\omega\, \omega^{3}\frac{\sin^{2}\left((\omega+\Omega)\frac{t}{2}\right)}{\left((\omega+\Omega)\frac{1}{2}\right)^{2}},\label{eq38}\\
P_{\text{min}}\rightarrow&\frac{262144\hbar^{3}\epsilon_{0}}{177147 c^{3}q^{2}Z^{2}\mu_{e}^{2}}\int\limits_{0}^{\omega\ll \frac{Z}{a_{0}}}\dd\omega \,\omega^{3}\frac{\sin^{2}\left((\omega+\Omega)\frac{t}{2}\right)}{\left((\omega+\Omega)\frac{1}{2}\right)^{2}},\label{eq39}
\end{align}
which, after substituting numerical values becomes
\begin{align}
    P_{\text{dip}},P_{\text{min}}\rightarrow&
2.68\times 10^{-41}s^{2}
\int\limits_{0}^{\omega\ll \frac{1}{a_{0}}}\dd\omega \,\omega^{3}\frac{\sin^{2}\left((\omega+\Omega)\frac{t}{2}\right)}{\left((\omega+\Omega)\frac{1}{2}\right)^{2}}.\label{eq40}
\end{align}

Under normal circumstances the dipole approximation criterion is not satisfied by higher frequency modes and we therefore ask how large are the contributions from these non-dipole approximation modes and how large is the subsequent difference between the models. This will identify scales for which the dipole approximation is accurate even if there is no dominant frequency. As we will discuss below, for example, this would include interactions where the atom and the field are in their ground states but the interaction lasts longer than the light-crossing time of the atom.

From \eqref{eq31} and \eqref{eq32} we cannot, a priori, know the exact effect of the high frequency modes on the transition probabilities. Our only expectation is that the introduction of higher order multipoles would reduce the discrepency between the two models, however that is not relevant to this manuscript. We re-emphasise that we want to assess the validity of this approximation in processes like vacuum excitations where there is no range of dominant frequencies and the duration of the interaction is what will dictate whether the approximation is good.  

\subsection{Excited fields}
When considering optical experiments, one of the most common excited fields considered is the coherent state. This is the state usually associated with a laser and for our purposes we will model it with a Gaussian frequency spectrum
\begin{align}
\ket{\phi_{i}}&=\mathcal{N}\exp\left(\int\dd\bm{k} \sum_{\lambda=1}^{2}G_{\lambda}(\bm{k},\bm{k}_{0})\hat{a}^{\dagger}_{\lambda}(\bm{k})\right)\ket{0},
\end{align}
where $G_{\lambda}(\bm{k},\bm{k}_{0})$ is some Gaussian centred at $\bm{k}_{0}$ and $\mathcal{N}$ is the appropriate normalisation factor. Coherent states are eigensates of the field annihilation operator, using this fact we can simplify \eqref{eq30} as
\begin{align}
\begin{split}
&P(i\rightarrow f)\\
&=\sum_{\lambda=1}^{2}\int\dd^{3}\bm{k}\left(h^{*}_{1,\lambda}(\bm{k})G_{\lambda}^{*}(\bm{k},\bm{k}_{0})+h^{*}_{2,\lambda}(\bm{k})G_{\lambda}^{\vphantom{*}}(\bm{k},\bm{k}_{0})\right)\\
\times&\sum_{\lambda=1}^{2}\int\dd^{3}\bm{k}' \left(h_{1,\lambda'}(\bm{k}')G_{\lambda'}^{\vphantom{*}}(\bm{k}',\bm{k}_{0})+h_{2,\lambda'}(\bm{k}')G_{\lambda'}^{*}(\bm{k}',\bm{k}_{0})\right)\\
+&\sum_{\lambda=1}^{2}\int\dd^{3}\bm{k}h_{2,\lambda}^{*}(\bm{k})h_{2,\lambda}(\bm{k}),
\end{split}\notag\\
\begin{split}
&=\abs{\sum_{\lambda=1}^{2}\int\dd^{3}\bm{k}\left(h^{*}_{1,\lambda}(\bm{k})G_{\lambda}^{*}(\bm{k})+h^{*}_{2,\lambda}(\bm{k})G_{\lambda}^{\vphantom{*}}(\bm{k})\right)}^{2}\\
+&\sum_{\lambda=1}^{2}\int\dd^{3}\bm{k}h_{2,\lambda}^{*}(\bm{k})h_{2,\lambda}(\bm{k}),
\end{split}
\end{align}

For the transition $1s\rightarrow 2p$ we have
\begin{align}
\begin{split}
h_{\pm,\lambda}^{\text{dip}}&=\pm
\frac{128\sqrt{2}\hbar^{3/2}\sqrt{\epsilon_{0}} t}{243\sqrt{\pi}q Z \mu_{e}}\\
&\frac{e^{-\frac{\ii t}{2}\left(2 E_{f}/\hbar\pm \omega-\Omega\right)}\text{sinc}\left(\frac{\omega\mp\Omega}{2}t\right)\sqrt{\omega}\sin(\theta_{k})}{\left(1+\frac{4 a_{0}^{2}}{9 c^{2}Z^{2}}\omega^{2}\right)^{3}},
\end{split}\\
\begin{split}
h_{\pm,\lambda}^{\text{min}}&=\pm\frac{128\sqrt{2}\hbar^{3/2}\sqrt{\epsilon_{0}} t}{243\sqrt{\pi}q Z \mu_{e}}\\
&\frac{e^{-\frac{\ii t}{2}\left(2 E_{f}/\hbar\pm \omega-\Omega\right)}\text{sinc}\left(\frac{\omega\mp\Omega}{2}t\right)\sqrt{\omega}\sin(\theta_{k})}{\left(1+\frac{4 a_{0}^{2}}{9 c^{2}Z^{2}}\omega^{2}\right)^{2}},
\end{split}
\end{align}
where the $\pm$ subscript refers to $+\rightarrow 1$, $-\rightarrow 2$; and $\theta_{k}$ is the spherical coordinate polar angle of $\bm{k}$. Here we have reintroduced the physical constants for completeness.

As with the vacuum contributions the minimal coupling and the dipole approximation differ only in the decay rate of $h_{\pm,\lambda}$ with respect to $\omega$. Unlike the vacuum contributions \eqref{eq31} and \eqref{eq32}, the asymptotic behaviour of $G_{\lambda}(\bm{k},\bm{k}_{0})$ make it possible to enforce the dipole approximation $\omega a_{0}/Z\ll 1$ for all significantly contributing modes. In this case we can implement a zeroth order Taylor expansion to obtain
\begin{align}
\begin{split}
h_{\pm,\lambda}^{\text{dip}}\rightarrow&\pm\frac{128\sqrt{2}\hbar^{3/2}\sqrt{\epsilon_{0}} t}{243\sqrt{\pi}q Z \mu_{e}}\\
&e^{-\frac{\ii t}{2}\left(2 E_{f}/\hbar\pm \omega-\Omega\right)}\text{sinc}\left(\frac{\omega\mp\Omega}{2}t\right)\sqrt{\omega}\sin(\theta_{k}),
\end{split}\\
\begin{split}
h_{\pm,\lambda}^{\text{min}}\rightarrow&\pm\frac{128\sqrt{2}\hbar^{3/2}\sqrt{\epsilon_{0}} t}{243\sqrt{\pi}q Z \mu_{e}}\\
&e^{-\frac{\ii t}{2}\left(2 E_{f}\pm \omega-\Omega\right)}\text{sinc}\left(\frac{\omega\mp\Omega}{2}t\right)\sqrt{\omega}\sin(\theta_{k}),
\end{split}
\end{align}
which are equal, as expected when the dipole approximation criterion is satisfied. 

Therefore, if the field dependent term of \eqref{eq30} is dominant over the vacuum contribution, and the field is excited in dipole approximation satisfying modes, then we expect that the dipole model can be successfully and reliably used.

\begin{figure}[!t]
\includegraphics[width=0.9\columnwidth]{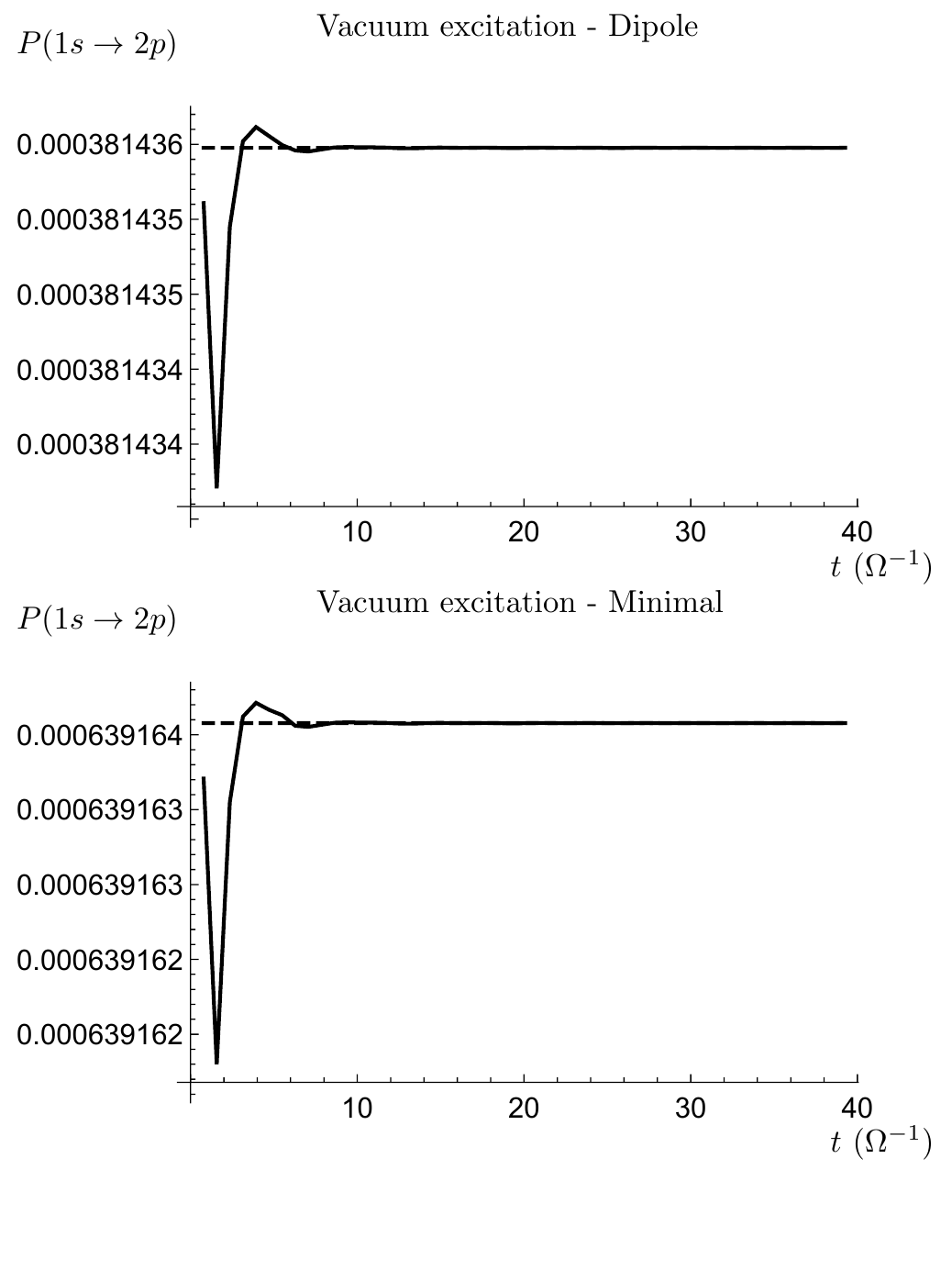}
\caption{Vacuum transition probability $1s\rightarrow 2p$ for $Z=1$ atom. The short time behaviour, i.e. $t\in(0,40\Omega^{-1})$, corresponds to all vacuum modes constructively contribution with very minor phase differences. In the long time limit all modes are dephased with one another, leading to a constant transition probability. The dashed lines correspond to analytic results obtained from \eqref{eq31} and \eqref{eq32} via $\begin{aligned}[t] \lim_{t\rightarrow\infty}\sin^{2}\left((\omega+\Omega)\frac{t}{2}\right)=\frac{1}{2}\end{aligned}$. From these analytic expressions we know that the offset is $2.58\times 10^{-4}$.}\label{fig1}
\end{figure}

\section{Results}
Given the nature of the integrals in question we resort to numerical integration in order to study the discrepancies between dipole and minimal models. In particular our interest lies in how the probability of transition varies with the size of the electron orbital. Scully and Zubairy \cite{Scully} seem to suggest that in the limit of an infinitely small atom the two models should converge; however that derivation was based around classical fields whilst assuming the energy gap $\Omega$ remains constant as the atom is shrunk.

\subsection{Vacuum fields}
\subsubsection{Vacuum excitation}
Consider first the transition $1s\rightarrow 2p$ with the initial EM field in the vacuum.

In Fig.~\ref{fig1} the transition probabilities have been plotted as a function of time. At a glance these two figures appear similar; however, the two graphs are offset by ~$2.6\times 10^{-4}$. Since the graphs do not detail extremely small times we must presume that this offset arises in the very early evolution of the electron, an artifact of using different models combined with a `sudden switching'. This already suggests that a significant difference is present for short time scales, i.e. $t<\Omega^{-1}$.

\begin{figure}[!t]
\includegraphics[width=0.9\columnwidth]{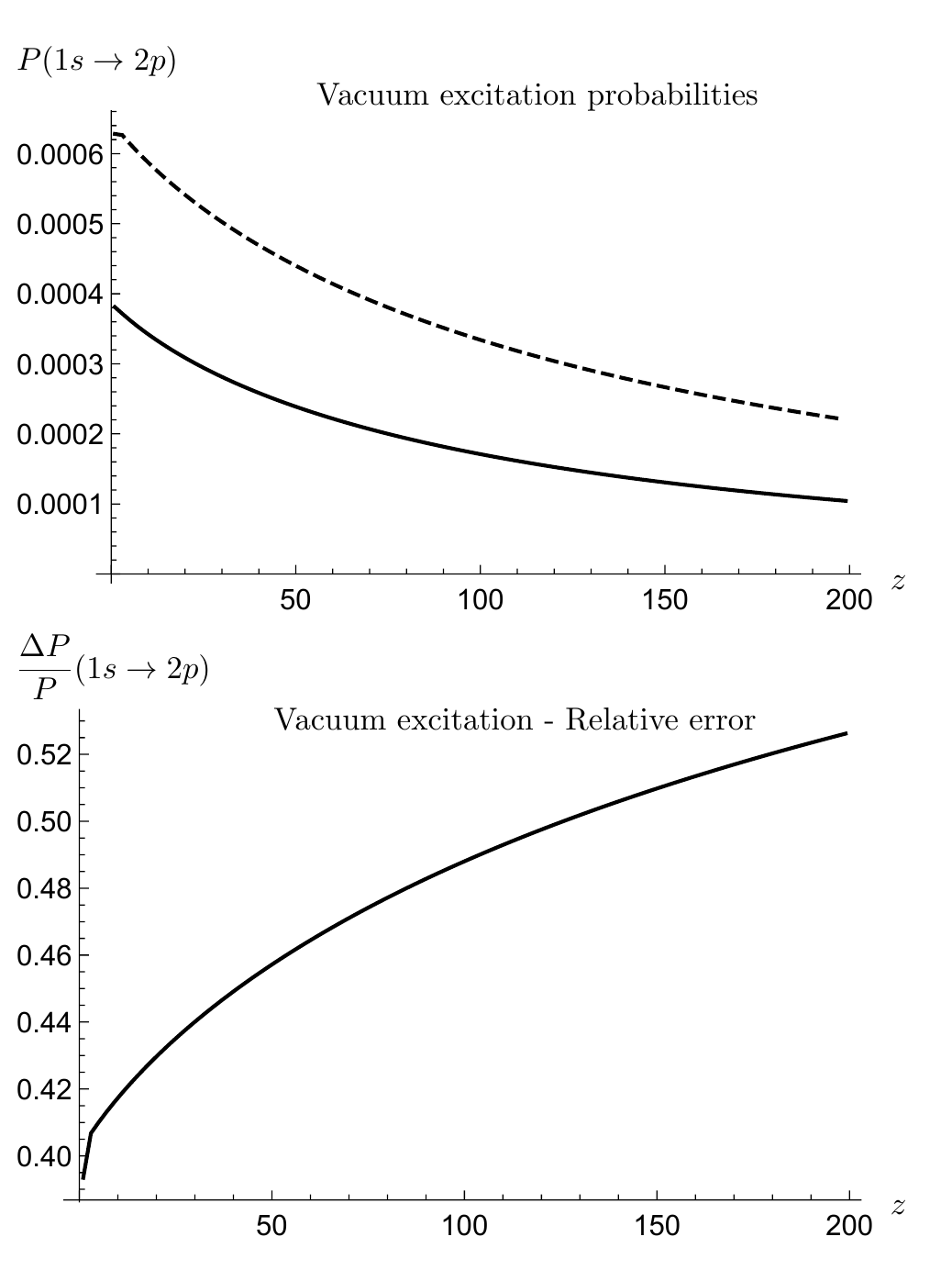}
\caption{Long time transition probability as a function of $Z$ and the relative error between the two models. As $Z$ increases the atom becomes smaller; however, contrary to Scully and Zubairy, the models diverge. Minimal model is dashed.}\label{fig2}
\end{figure}

In Fig.~\ref{fig2}, as expected, the transition probability decays with $Z$ given that the energy gap increases without an increase in the interaction strength. However in spite of the electron orbital size decreasing as $1/Z$ the predictions of the dipole and minimal model remain distant, in particular the relative error is also seen to increase.

This behaviour goes against our expectations given the dipole criterion $R\abs{\bm{k}}\ll 1$. Mathematically (see \eqref{eq31} and \eqref{eq32}) this behaviour is a consequence of the $\big(1+\frac{4 a_{0}^{2}}{9 c^{2}Z^{2}}\omega^{2}\big)$ term increasing the number of dipole approximation satisfying modes, whilst increasing the sensitivity to previously `dormant' UV modes via the $\omega^{3}\text{sinc}^{2}\left((\omega+\Omega)\frac{t}{2}\right)\sim \omega$ growth. These competing effects ensure that the two predictions never coincide.

\subsubsection{Spontanous emission}
Now consider the transition $2p\rightarrow 1s$ in the vacuum. In this case we would expect that if $\Omega a_{0}/Z\ll 1$ then the single mode approximation would limit the integration domain of \eqref{eq31} and \eqref{eq32} to a dipole approximation satisfying domain and therefore we expect the dipole model to be good.

\begin{figure}[!t]
\includegraphics[width=0.9\columnwidth]{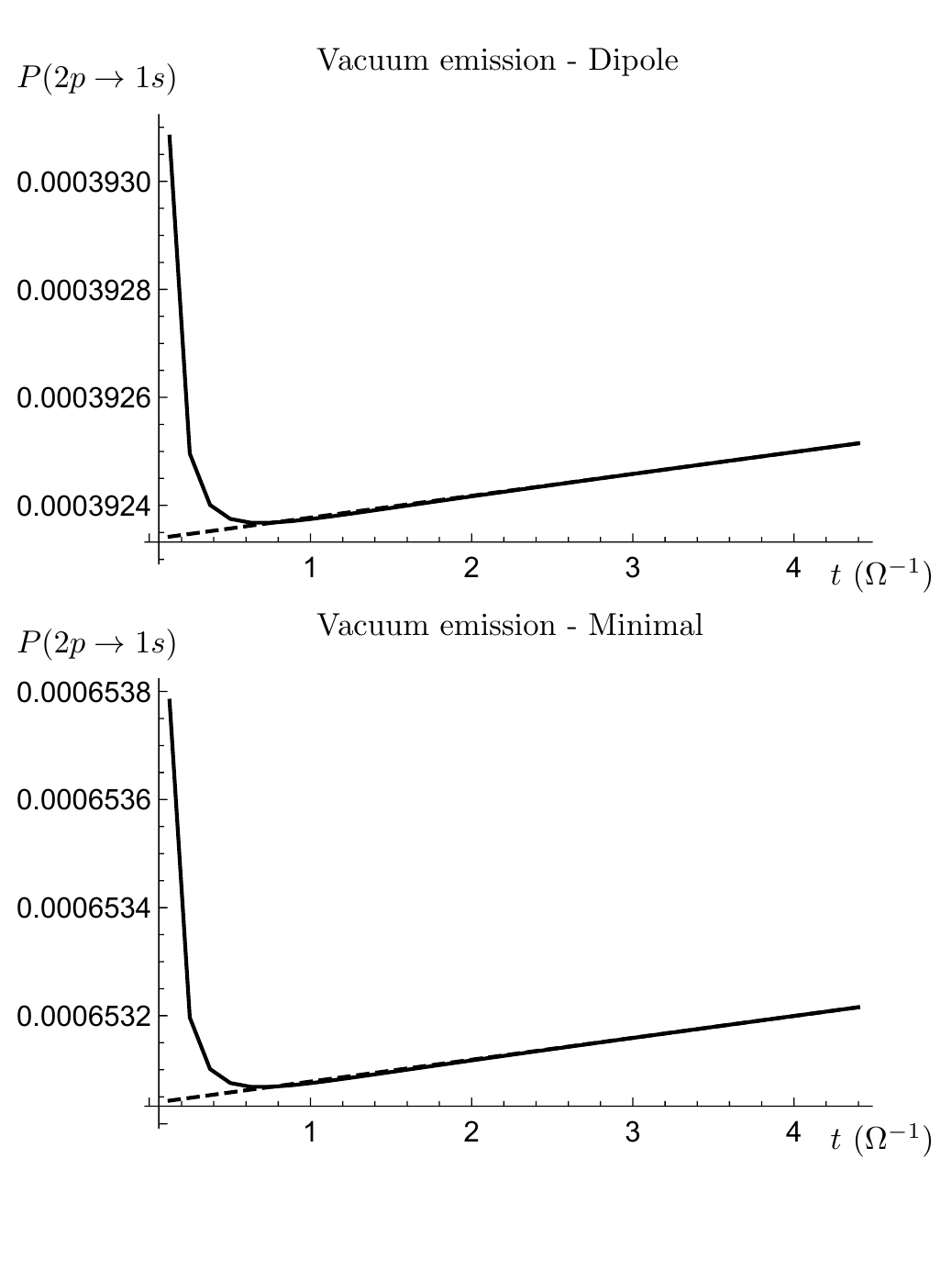}
\caption{Vacuum transition probability $2p\rightarrow 1s$ for $Z=1$ atom. The short time behaviour corresponds to the region where the single mode approximation is invalid. During this time the dipole approximation is violated and the observed offset is generated. The dashed line corresponds to the single mode approximation. Note that for longer times the curve becomes linear as dictated by the single mode approximation.}\label{fig5}
\end{figure}

In Fig.~\ref{fig5} the emission probabilities as a function of time are plotted. Their long time behaviour coincides with Fermi's golden rule, where both models yield the same gradient, implying coincidence of the models. This can be justified using the single mode approximation, which is valid for long times, i.e. 
\begin{align} \lim_{t\rightarrow\infty}t^{2}\text{sinc}^{2}\left((\omega+\Omega)\frac{t}{2}\right)\sim \pi t\delta\left(\frac{\omega+\Omega}{2}\right);\end{align} and the fact that $\Omega(Z=1)$ satisfies the dipole approximation. For short times the single mode approximation is no longer applicable and this generates the offset seen in the graphs.

\begin{figure}[!t]
\includegraphics[width=0.9\columnwidth]{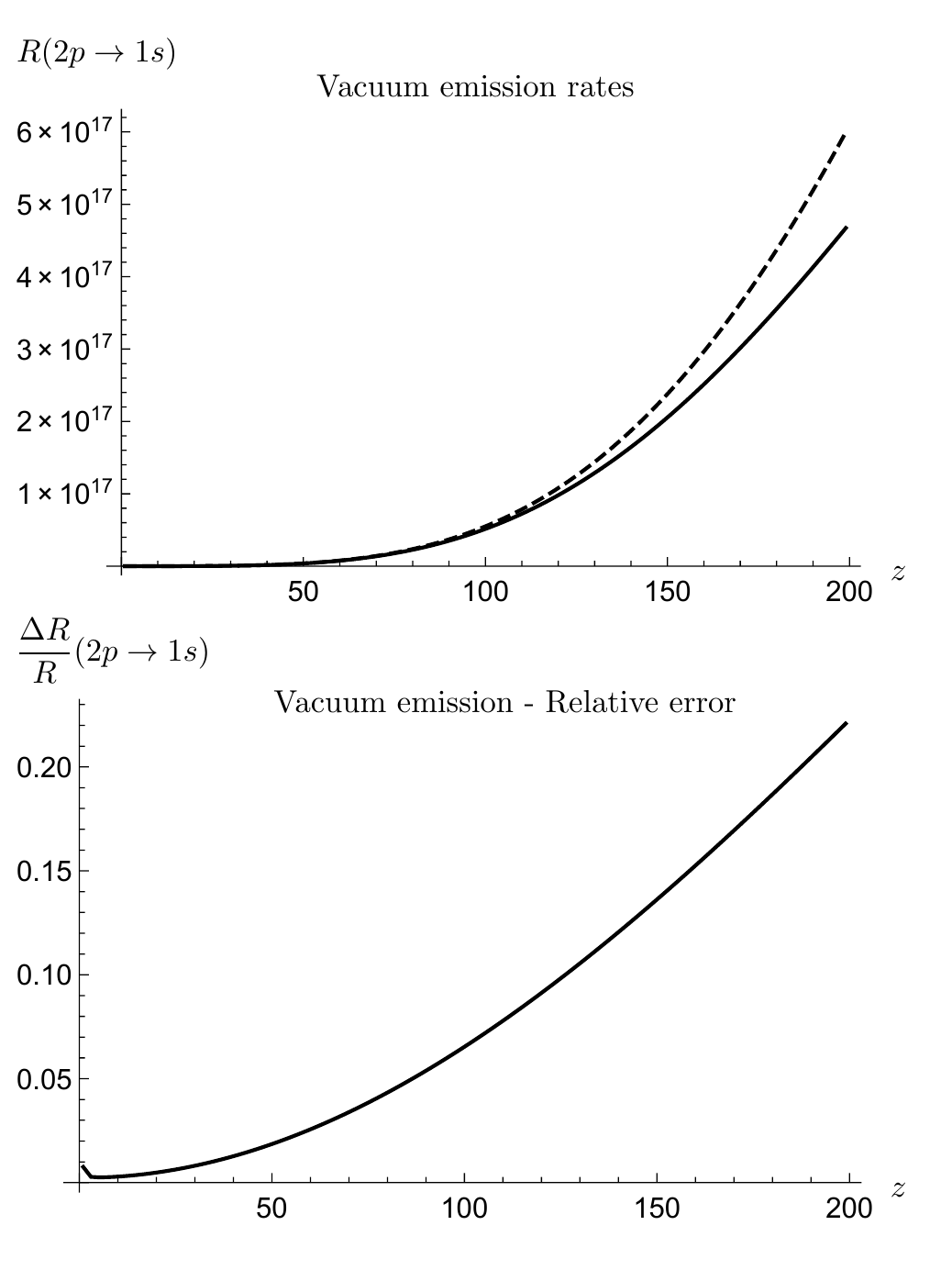}
\caption{Vacuum emission transition rates as a function of $Z$. Note that as $Z$ increases and the atom becomes smaller the two models diverge.  Minimal model is dashed. By implementing the single mode approximation in \eqref{eq31} and \eqref{eq32} one can show that the transition rates are given by $\begin{aligned}[t] R=\frac{6.26\times10^{8}Z^{4}}{\left(1+3.33\times 10^{-6}\right)^{n}}\end{aligned}$, where $n=4,6$ for the minimal and dipole coupling respectively.}\label{fig6}
\end{figure}

Fig.~\ref{fig6} shows the progression of the asymptotic emission rate with $Z$. As $Z$ increases the atom size decreases; however when implementing the single mode approximation the dipole approximation criterion becomes $\Omega a_{0}/Z\ll 1$, so the atomic size decreases as $1/Z$ but the energy gap $\Omega$ increases as $Z^{2}$, therefore as $Z\rightarrow \infty$ the dipole criterion is increasingly violated. Of interest is the relative error graph in Fig.~\ref{fig6}. The linear vs quadratic behaviour results in a minimum in the relative error occurring at $Z\approx 3$, the optimal proton number for coincidence of minimal and dipole models.

Note that if a temporal switching is introduced this may help the dipole model converge on the minimal model, at the cost of invalidating any use or interpretations of the single mode approximation.

\subsection{Excited fields}\label{sec4b}
Finally consider the transitions concerning excited fields, i.e. field states where $P_{\phi}$ from \eqref{eq30} is not zero. In particular we focus on `spatial pulses' of coherent `light'. In order to explore the effects of model choice on $P_{\phi}$ alone the following section will involve plots and discussions of $P_{\phi}$ alone, note that $P_{0}$ is independent of the field state so the discussion in previous sections generally holds.

\begin{figure}[!t]
\includegraphics[width=0.9\columnwidth]{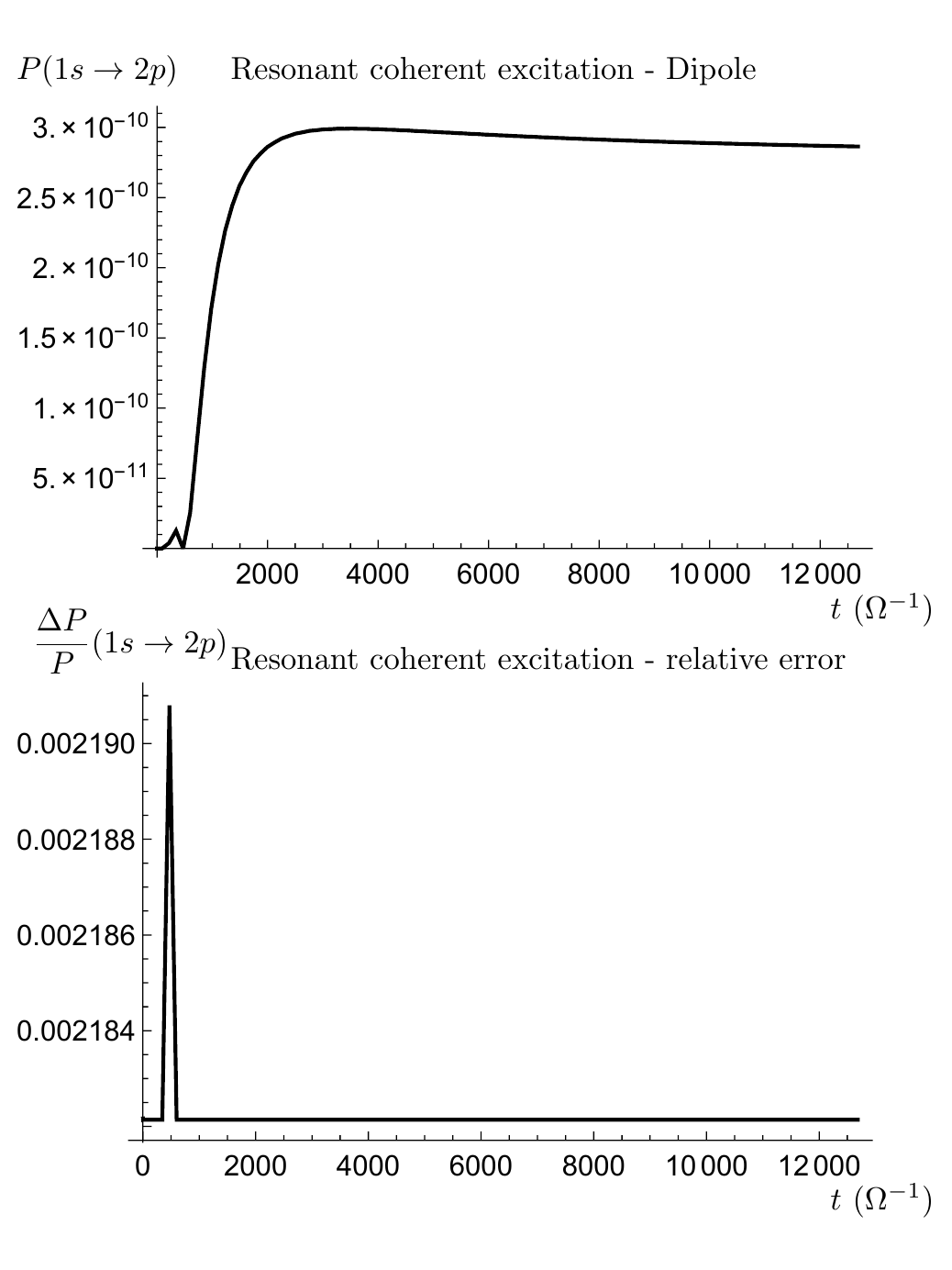}
\caption{Excitation probability ($P_{\phi}$ only) for a Gaussian coherent pulse with central frequency $\Omega$ (resonant). As the pulse arrives the transition probability increases to $3\times 10^{-10}$. Once the pulse is far from the atom the transition probability remains roughly constant. Note that the relative error remains small throughout ~$2\times 10^{-3}$. Also note, the bump at early times is believed to be a numerical imprecision.}\label{fig7}
\end{figure}

\begin{figure}[!t]
\includegraphics[width=0.9\columnwidth]{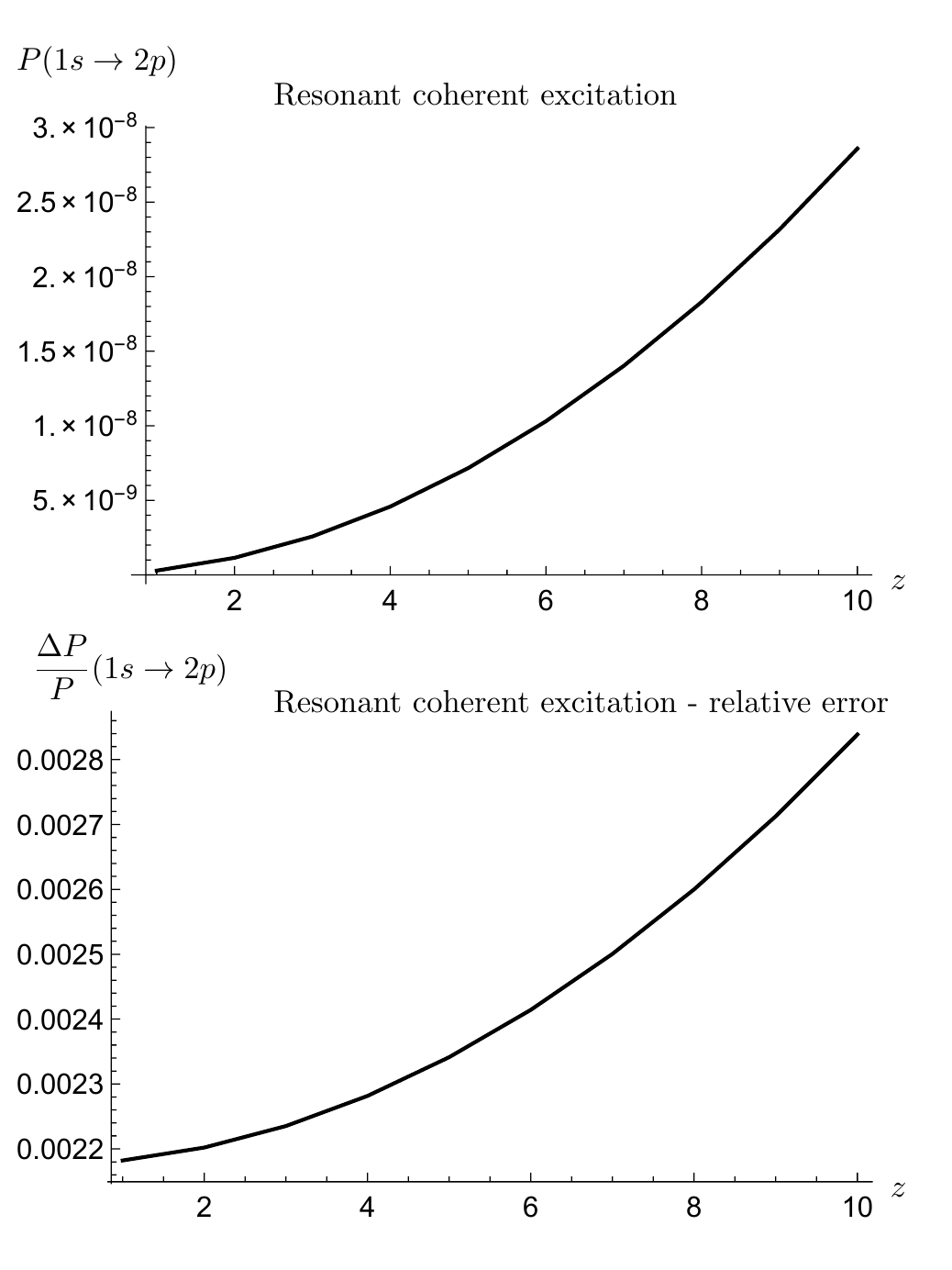}
\caption{Long time excitation probability ($P_{\phi}$ only) as a function of $Z$. Unlike the vacuum cases the dominant frequency is dictated by the field excitation and not the single mode approximation; however, since we chose the field excitation to remain resonant with the atomic transition the relative error increases with $Z$ as $\Omega$ ceases to satisfy the dipole approximation.}\label{fig8}
\end{figure}

The initial field state used was
\begin{align}
\ket{\phi}&=\mathcal{N}\exp\left(\sum_{\lambda}\int\dd^{3}\bm{k}\,G_{\lambda}(\bm{k})\hat{a}_{\lambda}^{\dagger}(\bm{k})\right)\ket{0},
\end{align}
where
\begin{align}
G_{\lambda}(\bm{k})&=\frac{\delta_{\lambda,\lambda_{0}}e^{\ii \omega T^{*}}}{(2\pi)^{3/4}}\frac{e^{-\frac{\left(k_{x}-k^{0}_{x}\right)^{2}}{4\sigma_{x}^{2}}}e^{-\frac{\left(k_{y}-k^{0}_{y}\right)^{2}}{4\sigma_{y}^{2}}}e^{-\frac{\left(k_{z}-k^{0}_{z}\right)^{2}}{4\sigma_{z}^{2}}}}{\sqrt{\sigma_{x}\sigma_{y}\sigma_{z}}},
\end{align}
where $\bm{k}^{0}$ is the central wavevector of the wavepacket, $\lambda_{0}$ is the polarisation of the field excitation, $T^{*}$ dictates the wavepackets initial position and $\mathcal{N}$ is the appropriate normalisation factor. Note that $G_{\lambda}(\bm{k})$ is $L_{2}$ normalised.

For the numerical work presented here $\sigma_{x}=\sigma_{y}=\sigma_{z}=\Omega/100$ and $\bm{k}^{0}=\Omega\bm{e}_{x}$, i.e. a resonant wavepacket. Fig.~\ref{fig7} shows the $P_{\phi}$ contribution to the transition probability as a function of time. There is a rapid increase in the transition probability as the wavepacket passes through the atom, finally the probability becomes almost constant as the field locally returns to the vacuum. The relative error between the two models is very small, in fact it is very similar to the relative error shown in Fig.~\ref{fig6} for small $Z$. Note that as we change $Z$ then $\Omega\sim Z^{2}$; this includes changing the EM field.

In Fig.~\ref{fig8} the asymptotic transition probability is shown along with the relative error between the models as a function of $Z$. As in previous cases as $Z$ increases the dominant mode $\Omega$ no longer satisfies the dipole approximation and therefore there is no expectation that the two models should give the same predictions.

\section{Discussion}
As can be seen from the plots above there seem to be cases when the dipole model is valid and others when it is not. These can be explained by the existence or not of a dominant mode and whether this mode satisfies the dipole approximation.

\begin{figure}[!t]
\includegraphics[width=0.9\columnwidth]{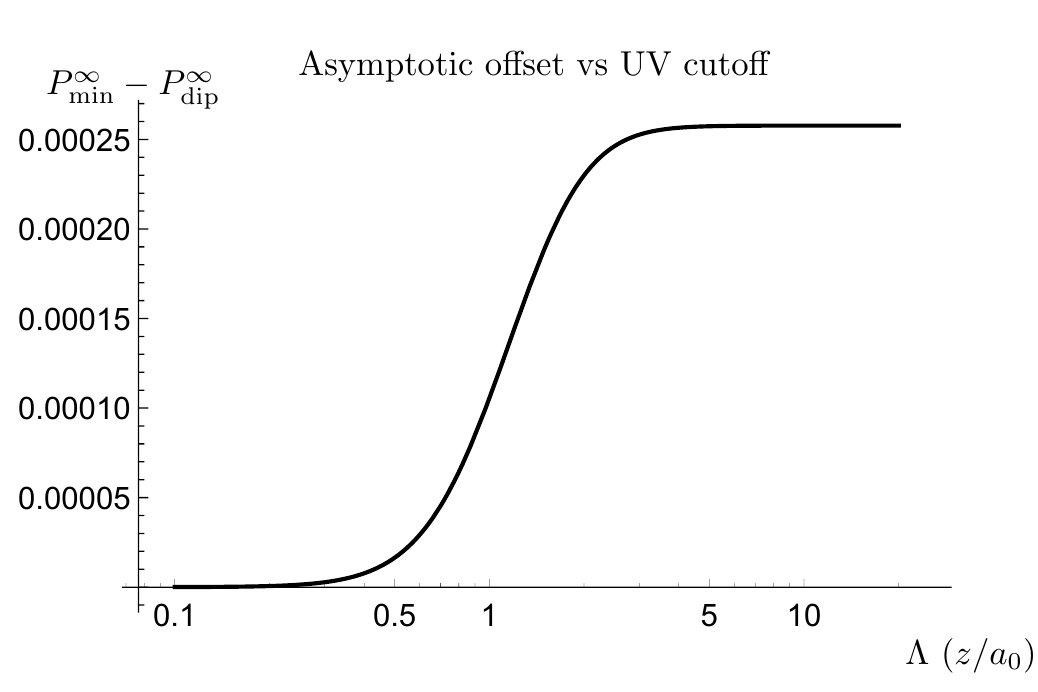}
\caption{Difference in asymptotic behaviour between dipole and minimal models for vacuum excitations, as a function of a hard UV cutoff. The x-axis is normalised to dipole approximation frequency cutoff. $\Lambda\ll 1$ corresponds to an application of the dipole approximation. This reflects the conclusion of \eqref{eq40}.}\label{fig9}
\end{figure}

In the case of vacuum excitations there is no notion of a dominant mode in the EM field. The vacuum fluctuations cause all modes of all wavelengths to interact with the electron, with short wavelength modes suppressed by the Fourier properties of the atom itself. In particular the equation describing the contributions of each mode is given by \eqref{eq31} and \eqref{eq32}. It can be rewritten to highlight key aspects as

\begin{align}
P&=K\int\dd\omega\underbrace{\frac{1}{(1+\frac{4 a_{0}^{2}\omega^{2}}{9Z^{2}})^{n}}}_{\text{Geometry \& coupling}}\underbrace{\omega^{3}\frac{\sin^{2}\left((\omega+\Omega)\frac{t}{2}\right)}{\left((\omega+\Omega)\frac{1}{2}\right)^{2}}}_{\text{Intrinsic \& Switching}},\label{eq45}
\end{align}
where $K$ is some constant and $n=4,6$ depending on the model. The intrinsic \& switching term dictates the `dominant' or `range of dominant' modes. When considering the dipole approximation $\omega\ll Z/a_{0}$ this can be interpreted as saying ``we want the intrinsic \& switching factors to decay long before the geometry \& coupling term begins to decay.'' As was shown in section \ref{sec3b} treating the geometry \& coupling term as constant reduces the minimal model to the dipole model. Inspection of \eqref{eq45} shows that the intrinsic \& switching term actually grows with increasing $\omega$, contrary to our needs; therefore creating this discrepancy between the two models. If a smooth switching function (with characteristic width $T$) could be introduced, then the intrinsic \& switching term would be modified to suppress UV modes with $\omega\gtrsim T^{-1}$, thereby reducing the contributions to the transition probability from high frequency modes and diminishing the difference between the dipole and minimal models. We illustrate this in figure \ref{fig9}, where we introduce a cutoff $\Lambda$ (that would be proportional to $1/T$) and we see that the two models yield identical predictions for small enough $\Lambda$. This characteristic appears in figure \ref{fig1}, where the high frequency modes cause an offset in the transition probabilities on a very short timescale. However for longer times the two models predict similar trends (i.e. for $T$ such that IR unsuppressed modes $\omega\lesssim T^{-1}$ satisfy the dipole criterion).

In the case of spontaneous emission $\Omega<0$ and so the intrinsic \& switching terms of \eqref{eq45} have a dominant frequency ($\omega=-\Omega$). It is the identity $\begin{aligned}[t] \lim_{t\rightarrow\infty}\frac{\sin^{2}(\eta t/2)}{(\eta/2)^{2}t}=\pi\delta(\eta/2)\end{aligned}$ that gives us the dominant mode, not by suppressing the higher frequency modes but by elevating a single mode, i.e. the single mode approximation. If this dominant mode satisfies the dipole approximation then the geometry \& coupling term will have the approximate value of $1$ and both models will be equivalent. However, if $\Omega$ no longer satisfies the dipole approximation then the models will begin to differ and this is what is shown in Fig.~\ref{fig5}. As $Z$ becomes larger, $\Omega$ grows quadratically in $Z$ and therefore at some point $\Omega>Z/a_{0}$.

In section \ref{sec4b} the excited field contributions are shown, demonstrating how similar the predictions of the two models are when there is a dominant mode that satisfies the dipole approximation. In particular Fig.~\ref{fig8} shows how the models begin to diverge as $\Omega$ ceases to satisfy the dipole approximation. These are the effective predictions given under the rotating wave approximation and they hold true for stronger coherent amplitudes. This is the regime most commonly found in experiments and therefore justifies the widespread use of the dipole model.

In the case of spontaneous emission or stimulated excitation we say that the dipole model is good because the dominant frequency $\Omega$ satisfies the dipole approximation; however, there is still a relative error of ~$2\times10^{-3}$. If we consider the basis of the dipole approximation, i.e. approximating $e^{\ii\bm{k}\cdot\bm{x}}\approx 1+\ii \bm{k}\cdot\bm{x}\approx 1$ then we note that the first order error will be of $O(\bm{k}\cdot\bm{x})$, which can be rewritten as $O(\Omega a_{0}/Z)$. Therefore the amount by which the dipole approximation is satisfied provides an estimate for the error between the two models. In particular note that for the hydrogen atom $\Omega a_{0}=2.7\times 10^{-3}$. In the cases where a dominant frequency exists, or a finite range of effective modes exists, the dipole criterion can be used as a first order estimate for the relative error. Hence if, experiementally, $a_{0}$ can be made smaller without changing $\Omega$ then the dipole model would become exact for all cases except for the vacuum excitation case, where the probability of transition would become divergent.

\section{Conclusion}

In this paper we have evaluated the differences in the predictions of the minimal coupling Hamiltonian ($\hat{\bm{p}}\cdot\hat{\bm{A}}$) and the dipole coupling ($\hat{\bm{x}}\cdot\hat{\bm{E}}$) of light-matter interaction for the hydrogen atom. This comparison is non-trivial for extended atoms due to the explicit gauge noninvariant nature of the minimal coupling.  

We have confirmed the validity of  the predictions of the dipole model for spontaneous emission and stimulated excitation (situations where there is a dominant contribution from a particular frequency scale satisfying  a dipole approximation) in the long time regimes. In these situations the relative difference between the models is approximately given (to first order) by $\Omega a_{0}$, i.e. the magnitude evaluated to coarsely  assess the validity of the  dipole approximation.

Crucially, we have found the situation to be much different in the case of vacuum excitation, where the atom starts in the ground state and the field in the vacuum. For the cases of vacuum excitation, a dominant mode is absent, and even considering a very small (point-like) atom does not guarantee that the dipole approximation is accurate, and there can be indeed a discrepancy between dipole and minimal model predictions. One cannot get away in this case just saying that the atom is `small', since there is no characteristic field wavelength dominating the interaction to compare it with. Entanglement harvesting and the Fermi problem are two such scenarios where one considers finite-time evolution  of the ground state of atoms and the field vacuum, and one may then need to be further justified to use the dipole approximation.

For this case, we have characterized the regimes where the dipole model does not suffice to predict the physics of the light-matter interaction. In particular we found that when considering vacuum excitations for short time interactions the dipole coupling does not yield the same results that the minimal coupling, which is of particular interest when analyzing vacuum phenomena.

As \eqref{eq45} shows, this difference cannot be removed by shrinking the atom, and we have shown that it is the contribution of arbitrarily high frequency modes what makes the two predictions diverge.

However, in practice, most light-matter interactions are finite in time (preparation to measurement). We have shown that the introduction of a smooth switching (which in turn suppresses the influence of the higher frequency modes on the atomic dynamics) ensures satisfaction of the dipole criterion as long as the interaction time between atom and field is longer than the light-crossing time of the radius of the atom. This would justify the widespread use of the dipole approximation in modelling light-matter interactions even for vacuum fluctuations (as in the case of entanglement harvesting \cite{Reznik,VALENTINI1991321,Pozas2015} and the Fermi problem \cite{RevModPhys.4.87,PhysRevLett.107.150402}), but crucially not because of an argument of a `small atom', but instead for `sufficiently long interaction time'.

\acknowledgments

J.L. thanks the Institute for Quantum Computing at the University of Waterloo for hospitality. JL was supported in part by Science and Technology Facilities Council (Theory Consolidated Grant ST/P000703/1). E.M-M acknowledges support of the NSERC Discovery program as well as his Ontario Early Researcher Award.

\appendix 
\begin{widetext}

\section{Gauge transformations}\label{seca0}

If we begin by analysing Schr\"{o}dinger's equation we have
\begin{align}
\ii\frac{\partial\psi(\bm{x},t)}{\partial t}&=\left\{\frac{1}{2\mu_{e}}\left(\hat{p}-q\bm{A}(\bm{x},t)\right)^{2}+V(\bm{x})+q U(\bm{x},t)\right\}\psi(\bm{x},t).
\end{align}
If we introduce a local phase to the wavefunction of the form
\begin{align}
    \tilde{\psi}(\bm{x},t)= e^{-\ii q \chi(\bm{x},t)}\psi(\bm{x},t)
\end{align}
this will lead to a new Schr\"{o}dinger's equation:
\begin{align}
\ii\frac{\partial}{\partial t}\tilde{\psi}(\bm{x},t)=\left\{\frac{1}{2\mu_{e}}\left(\hat{p}-q\left[\bm{A}(\bm{x},t)-\nabla \chi\right]\right)^{2}+V(\bm{x})+q(U(\bm{x},t)+\dot{\chi})\right\}\tilde{\psi}(\bm{x},t).
\end{align}
This new equation is also Schr\"{o}dinger's equation for a gauge transformed EM field, i.e.
\begin{align}
\bm{A}(\bm{x},t)\rightarrow&\bm{A}(\bm{x},t)-\nabla\chi,\\
U(\bm{x},t)\rightarrow& U(\bm{x},t)+\dot{\chi}.
\end{align}

Now if we start with Schr\"{o}dinger's equation in the Coulomb gauge
\begin{align}
U&=0,\\
\nabla\cdot\bm{A}&=0,
\end{align}
and then perform the gauge transformation
\begin{align}
\chi=\bm{A}(\bm{x},t)\cdot\bm{x},
\end{align}
this will yield an equation
\begin{align}
\begin{split}
i\frac{\partial\tilde{\psi}(\bm{x},t)}{\partial t}
&=\left\{\frac{1}{2\mu_{e}}\left(\hat{\bm{p}}+q\left[\left(x_{i}\nabla\right)A_{i}(\bm{x},t)\right]\right)^{2}-q\hat{\bm{x}}\cdot\bm{E}(\bm{x},t)+V(r)\right\}\tilde{\psi}(\bm{x},t).
\end{split}
\end{align}
Note that under the conditions of the dipole approximation $\bm{A}$ is considered constant over the support of the wavefunction, hence the term
$\left[\left(x_{i}\nabla\right)A_{i}(\bm{x},t)\right]$ will be zero, exactly giving the dipole approximation.

This gauge transformation was chosen specifically because for small atoms (with respect to some EM wavelength) this will reproduce the dipole approximation, which exactly the approximation we wish to test for different parameter regimes.

\section{Gauge invariance of transition probabilities between dressed states}\label{seca05}

Consider the minimal model Hamiltonian
\begin{align}
\hat{H}&=\frac{1}{2\mu_{e}}\left(\hat{\bm{p}}-e\bm{A}(\hat{\bm{x}},t)\right)^{2}+V(\hat{\bm{x}})+qU(\hat{\bm{x}},t)\notag\\
&=\hat{H}_{0}+q\hat{H}_{1}+O(q^{2}),
\end{align}
where we discard $O(q^{2})$ terms since we shall be performing a first order perturbative expansion. 
To first order the dressed state can be written as
\begin{align}
\ket{\tilde{\psi}_{t,l}}=\sum_{k}\left(\delta_{lk}+\ii q L_{lk}(t)\right)\ket{\psi_{k}},
\end{align}
where $\ket{\psi_{k}}$ are the $E_{k}$ eigenstates of $\hat{H}_{0}$. These form the measurement basis for the electron in the particular EM gauge we choose. 

Since we work to first order perturbation theory, the evolving state of the electron can be expressed as
\begin{align}
\ket{\tilde{\psi}_{l}(t)}=\sum_{k}\left(\delta_{lk}+\ii q K_{lk}(t)\right)e^{-\ii E_{k}t}\ket{\psi_{k}},
\end{align}
where the initial condition is $K_{lk}(0)=L_{lk}(0)$. In order to determine the time evolution of $K_{lk}(t)$ we need to use Schr\"{o}dinger's equation,
\begin{align}
\ii\frac{\partial}{\partial t}\ket{\tilde{\psi}_{l}(t)}&=\sum_{k}\left(\delta_{lk}+\ii qK_{lk}(t)\right) E_{k}e^{-\ii E_{k}t}\ket{\psi_{k}}
-\sum_{k}q\dot{K}_{lk}(t)e^{-\ii E_{k}t}\ket{\psi_{k}}\notag\\
&=\hat{H}_{0}\sum_{k}\left(\delta_{lk}+\ii q K_{lk}(t)\right)e^{-\ii E_{k}t}\ket{\psi_{k}}+q\hat{H}_{1}\sum_{k}\delta_{lk}e^{-\ii E_{k}t}\ket{\psi_{k}}+O(q^{2}).
\end{align}
By cancelling out the appropriate terms and taking an inner product with $\bra{\psi_{m}}$ we are left with
\begin{align}
\dot{K}_{lm}(t)&=-\braket{\psi_{m}|\hat{H}_{1}|\psi_{l}}e^{\ii (E_{m}-E_{l})t},
\end{align}
and therefore
\begin{align}
K_{lk}(T)&=-\int\limits_{0}^{T}\dd t\braket{\psi_{k}|\hat{H}_{1}|\psi_{l}}e^{\ii (E_{k}-E_{l})t}+L_{lk}(0).
\end{align}

In the main text we made the choice
\begin{align}
L_{lk}(t)=\begin{cases}
-\frac{\braket{\psi_{k}|\frac{\bm{A}(\hat{\bm{x}},t)\cdot\hat{\bm{p}}+\hat{\bm{p}}\cdot\bm{A}(\hat{\bm{x}},t)}{2\mu_{e}}|\psi_{l}}}{\ii(E_{l}-E_{k})} & \text{if $E_{l}\neq E_{k}$},\\
-\bra{\psi_{k}}\int\limits_{0}^{t}\dd s\, U(\hat{\bm{x}},s)\ket{\psi_{l}} &\text{if $E_{l}=E_{k}$},
\end{cases}\label{eqb8}
\end{align}
and stated that this will guarantee gauge invariance of the transition amplitudes, and hence gauge invariance of the measurable transition probabilities. To prove this, consider first the following inner product,
\begin{align}
\braket{\psi_{k}|\hat{\bm{p}}\cdot\left(\nabla\chi\right)+\left(\nabla\chi\right)\cdot\hat{\bm{p}}|\psi_{l}}&=
\braket{\psi_{k}|\left(\hat{\bm{p}}\cdot\nabla\right)\chi-\hat{\bm{p}}\cdot\chi\nabla+\left(\nabla\chi\right)\cdot\hat{\bm{p}}|\psi_{l}}\notag\\
&=\braket{\psi_{k}|\left(\hat{\bm{p}}\cdot\nabla\right)\chi-\left(-\ii\nabla\chi\right)\cdot\nabla-\chi\hat{\bm{p}}\cdot\nabla+\left(\nabla\chi\right)\cdot\hat{\bm{p}}|\psi_{l}}\notag\\
&=\braket{\psi_{k}|\left(\hat{\bm{p}}\cdot\nabla\right)\chi-\left(\nabla\chi\right)\cdot\hat{\bm{p}}-\chi\hat{\bm{p}}\cdot\nabla+\left(\nabla\chi\right)\cdot\hat{\bm{p}}|\psi_{l}}\notag\\
&=\braket{\psi_{k}|\left(\hat{\bm{p}}\cdot\nabla\right)\chi-\chi\hat{\bm{p}}\cdot\nabla|\psi_{l}}\notag\\
&=\ii\braket{\psi_{k}|\hat{\bm{p}}^{2}\chi-\chi\hat{\bm{p}}^{2}|\psi_{l}},
\end{align}
where we have repeatedly implemented the product rule and the definition of the momentum operator. One can then exploit the fact that $\hat{H}_{0}\ket{\psi_{k}}=E_{k}\ket{\psi_{k}}$:
\begin{align}
\braket{\psi_{k}|\frac{\hat{\bm{p}}\cdot\left(\nabla\chi\right)+\left(\nabla\chi\right)\cdot\hat{\bm{p}}}{2\mu_{e}}|\psi_{l}}&=\ii\braket{\psi_{k}|\frac{\hat{\bm{p}}^{2}}{2\mu_{e}}\chi-\chi\frac{\hat{\bm{p}}^{2}}{2\mu_{e}}|\psi_{l}}\notag\\
&=\ii\braket{\psi_{k}|\left(\frac{\hat{\bm{p}}^{2}}{2\mu_{e}}+V(\hat{\bm{x}})\right)\chi-\chi\left(\frac{\hat{\bm{p}}^{2}}{2\mu_{e}}+V(\hat{\bm{x}})\right)|\psi_{l}}\notag\\
&=\ii\braket{\psi_{k}|\hat{H}_{0}\chi-\chi\hat{H}_{0}|\psi_{l}}\notag\\
&=\ii(E_{k}-E_{l})\braket{\psi_{k}|\chi|\psi_{l}}.
\end{align}
Armed with this identity consider $L_{lk}(t)\rightarrow L'_{lk}(t)$ under field gauge transformations. For $E_{l}\neq E_{k}$, we have 
\begin{align}
L'_{lk}(t)&=\frac{\braket{\psi_{k}|-\frac{\hat{\bm{p}}\cdot\bm{A}'(\hat{\bm{x}},t)+\bm{A}'(\hat{\bm{x}},t)\cdot\hat{\bm{p}}}{2\mu_{e}}|\psi_{l}}}{\ii(E_{l}-E_{k})}\notag\\
&=\frac{\braket{\psi_{k}|-\frac{\hat{\bm{p}}\cdot\bm{A}(\hat{\bm{x}},t)+\bm{A}(\hat{\bm{x}},t)\cdot\hat{\bm{p}}}{2\mu_{e}}|\psi_{l}}}{\ii(E_{l}-E_{k})}+\frac{\braket{\psi_{k}|
\frac{\hat{\bm{p}}\cdot\left(\nabla\chi\right)+\left(\nabla\chi\right)\cdot\hat{\bm{p}}}{2\mu_{e}}
|\psi_{l}}}{\ii(E_{l}-E_{k})}\notag\\
&=\frac{\braket{\psi_{k}|-\frac{\hat{\bm{p}}\cdot\bm{A}(\hat{\bm{x}},t)+\bm{A}(\hat{\bm{x}},t)\cdot\hat{\bm{p}}}{2\mu_{e}}|\psi_{l}}}{\ii(E_{l}-E_{k})}+\frac{\ii(E_{k}-E_{l})\braket{\psi_{k}|\chi|\psi_{l}}}{\ii(E_{l}-E_{k})}\notag\\
&=L_{lk}(t)-\braket{\psi_{k}|\chi(\hat{\bm{x}},t)|\psi_{l}}.
\end{align}
Similarly, for $E_{l}=E_{k}$, we have
\begin{align}
L_{lk}'(t)&=-\braket{\psi_{k}|\int\limits_{0}^{t}\dd s\, U'(\hat{\bm{x}},s)|\psi_{l}}\notag\\
&=-\braket{\psi_{k}|\int\limits_{0}^{t}\dd s\left( U(\hat{\bm{x}},s)+\dot{\chi}\right)|\psi_{l}}\notag\\
&=-\braket{\psi_{k}|\int\limits_{0}^{t}\dd s\, U(\hat{\bm{x}},s)|\psi_{l}}-\braket{\psi_{k}|\chi(\hat{\bm{x}},t)|\psi_{l}}+\braket{\psi_{k}|\chi(\hat{\bm{x}},0)|\psi_{l}}\notag\\
&=L_{lk}(t)-\braket{\psi_{k}|\chi(\hat{\bm{x}},t)|\psi_{l}}+\braket{\psi_{k}|\chi(\hat{\bm{x}},0)|\psi_{l}}.\label{eqb26}
\end{align}



Now consider the gauge transformation on $K_{lk}$,
\begin{align}
K'_{lk}(T)&=-\int\limits_{0}^{T}\dd t\braket{\psi_{k}|\hat{H}'_{1}|\psi_{l}}e^{\ii(E_{k}-E_{l})t}+L'_{lk}(0)\notag\\
&=-\int\limits_{0}^{T}\dd t\braket{\psi_{k}|\hat{H}_{1}+\frac{\hat{\bm{p}}\cdot\left(\nabla\chi\right)+\left(\nabla\chi\right)\cdot\hat{\bm{p}}}{2\mu_{e}}+\dot{\chi}|\psi_{l}}e^{\ii(E_{k}-E_{l})t}+L_{lk}(0)-\braket{\psi_{k}|\chi(\hat{\bm{x}},0)|\psi_{l}}\notag\\
&=-\int\limits_{0}^{T}\dd t\braket{\psi_{k}|\hat{H}_{1}|\psi_{l}} e^{\ii(E_{k}-E_{l})t}-\int\limits_{0}^{T}\dd t\braket{\psi_{k}|\ii (E_{k}-E_{l})\chi e^{\ii(E_{k}-E_{l})t}+\dot{\chi}e^{\ii(E_{k}-E_{l})t}|\psi_{l}}+L_{lk}(0)-\braket{\psi_{k}|\chi(\hat{\bm{x}},0)|\psi_{l}}\notag\\
&=K_{lk}(T)-\int\limits_{0}^{T}\dd t\frac{\dd}{\dd t}\braket{\psi_{k}|\chi e^{\ii(E_{k}-E_{l})t}|\psi_{l}}-\braket{\psi_{k}|\chi(\hat{\bm{x}},0)|\psi_{l}}\notag\\
&=K_{lk}(T)-\braket{\psi_{k}|\chi(\hat{\bm{x}},T)|\psi_{l}}e^{\ii(E_{k}-E_{l})T}.
\end{align}

As such, when considering measurement probability amplitudes ($l\neq k$),
\begin{align}
\braket{\tilde{\psi}_{T,k}|\tilde{\psi}_{l}(T)}&=\sum_{m}\bra{\psi_{m}}\left(\delta_{km}-\ii q L_{km}^{*}(T)\right)\sum_{n}\left(\delta_{ln}+\ii q K_{ln}\right)e^{-\ii E_{n}t}\ket{\psi_{n}}\notag\\
&=\ii q K_{lk}e^{-\ii E_{k}T}-\ii qL^{*}_{kl}e^{-\ii E_{l}T}+O(q^{2})\notag\\
&=\ii q\left(K_{lk}(T)e^{-\ii E_{k}T}-L^{*}_{kl}(T)e^{-\ii E_{l}T}\right)+O(q^{2}).
\end{align}
The gauge transformation properties of 
\begin{align}
K'_{lk}(T)e^{-\ii E_{k}T}-L'^{*}_{kl}(T)e^{-\ii E_{l}T}&=K_{lk}(T)e^{-\ii E_{k}T}-\braket{\psi_{k}|\chi(\hat{\bm{x}},T)|\psi_{l}}e^{-\ii E_{l}T}-L^{*}_{kl}(T)e^{-\ii E_{l}T}+\braket{\psi_{k}|\chi(\hat{\bm{x}},T)|\psi_{l}}e^{-\ii E_{l}T}\notag\\
&=K_{lk}(T)e^{-\ii E_{k}T}-L_{kl}^{*}(T)e^{-\ii E_{l}T},
\end{align}
therefore with our definition of $L_{lk}$ we now have gauge invariant transition amplitudes. In order to see how this has changed from the usual na\"{i}ve approach, repeat the process above with $L_{lk}=\delta_{lk}$ only.

\section{Perturbative time evolution of full quantum model}\label{seca2}
The models under scrutiny are the dipole model, computing the transition probability between states $\ket{1s}\rightarrow\ket{2p}$ under the Hamiltonian (to first order perturbation theory)
\begin{align}
\hat{H}=\hat{H}_{0}-q\hat{\bm{r}}\cdot\bm{E};
\end{align}
and the minimal model, which in the Coulomb gauge (where $\hat{\bm{p}}$ and $\hat{\bm{A}}$ commute)
\begin{align}
\hat{H}=\hat{H}_{0}-\frac{q}{\mu_{e}}\bm{A}\cdot\hat{\bm{p}}.
\end{align}
For completeness:
\begin{align}
\begin{split}
\hat{\bm{A}}(\bm{x},t)&=\int\frac{\dd^{3}\bm{k}}{(2\pi)^{3/2}\sqrt{2\omega}}\sum_{\lambda=1}^{2} \bm{\epsilon}_{\lambda}(\bm{k})\left(\hat{a}_{\lambda}^{\vphantom{\dagger}} (\bm{k})e^{-\ii(\omega t-\bm{k}\cdot\bm{x})}+\hat{a}_{\lambda}^{\dagger}(\bm{k})e^{\ii(\omega t-\bm{k}\cdot\bm{x})}\right),
\end{split}\\
\bm{k}\cdot\bm{\epsilon}_{\lambda}(\bm{k})&=0,\\
\hat{\bm{E}}(\bm{x},t)&=-\frac{\partial}{\partial t}\hat{\bm{A}}(\bm{x},t),
\end{align}

\subsection{Initial condition and measurement (aka dressed states)}
Taking the lead from the semi-classical derivation in appendix \ref{seca05}, the dressed states are
\begin{align}
\ket{\tilde{\psi}_{t,l},\phi_{i}}=\sum_{k}\left(\delta_{lk}+\ii q \hat{L}_{lk}(t)\right)\ket{\psi_{k}}\ket{\phi_{i}},
\end{align}
where
\begin{align}
\hat{L}_{lk}^{\text{min}}&=
\begin{cases}
-\frac{\braket{\psi_{k}|\frac{\hat{\bm{A}}(t)\cdot\hat{\bm{p}}}{\mu_e}|\psi_{l}}}{\ii(E_{l}-E_{k})} & \text{if $E_{l}\neq E_{k}$},\\
0 &\text{if $E_{l}=E_{k}$},
\end{cases}\\
\hat{L}_{lk}^{\text{dipole}}&=0,
\end{align}
and $\ket{\phi_{i}}$ is the initial state of the EM field. Note that $\hat{L}_{lk}$ are now operators acting on the EM field's Hilbert space. Therefore the initial state will be $\ket{\tilde{\psi}_{t,i},\phi_{i}}$ and the final measurement will involve an inner product of the time evolved state with $\ket{\tilde{\psi}_{t,f},\phi_{f}}$. 

Furthermore for this derivation
\begin{align}
\hat{H}&=\hat{H}_{0}+q\hat{H}_{1},\\
\hat{H}_{1}^{\text{min}}&=-\frac{\hat{\bm{A}}\cdot\hat{\bm{p}}}{\mu_{e}},\\
\hat{H}_{1}^{\text{dipole}}&=-\hat{\bm{E}}\cdot\hat{\bm{r}}.
\end{align}

As shown in appendix \ref{seca05} these definitions ensure the final measurement is gauge invariant (appendix \ref{seca05}'s results are independent of classical or quantum fields as they only require the gauge transformation properties of the EM potentials).

\subsection{Dynamics}
Given that we are looking at first order perturbations, the general state will be of the form
\begin{align}
\ket{\tilde{\psi}_{l}(t),\phi_{i}}&=\sum_{k}\left(\delta_{lk}+\ii q \hat{K}_{lk}(t)\right)e^{-\ii E_{k}t}\ket{\psi_{k}}\ket{\phi_{i}},
\end{align}
with initial conditions $\hat{K}_{lk}(0)=\hat{L}_{lk}(0)$. Recall that now $\hat{K}_{lk}$ is an operator acting on the EM field's Hilbert space. Schr\"{o}dinger's equation then becomes (EM field evolution has been encoded into the field operators via interaction picture)
\begin{align}
\ii\partial_{t}\ket{\tilde{\psi}_{l}(t),\phi_{i}}&=\sum_{k}\left(\delta_{lk}+\ii q \hat{K}_{lk}(t)\right)E_{k}e^{-\ii E_{k}t} \ket{\psi_{k}}\ket{\phi_{i}}+q\sum_{k}-\dot{\hat{K}}_{lk}(t)e^{-\ii E_{k}t}\ket{\psi_{k}}\ket{\phi_{i}}\notag\\
&=\left(\hat{H}_{0}+q\hat{H}_{1}\right)\ket{\tilde{\psi}_{l}(t)}\ket{\phi_{i}}\notag\\
&=\sum_{k}\left(\delta_{lk}+\ii q \hat{K}_{lk}(t)\right)E_{k}e^{-\ii E_{k}t} \ket{\psi_{k}}\ket{\phi_{i}}+q\hat{H}_{1}e^{-\ii E_{l}t}\ket{\psi_{l}}\ket{\phi_{i}}+O(q^{2}).
\end{align}
This leaves
\begin{align}
\dot{\hat{K}}_{lk}(t)&=-\braket{\psi_{k}|\hat{H}_{1}|\psi_{l}}e^{\ii(E_{k}-E_{l})t}.
\end{align}
Integration over time yields
\begin{align}
\hat{K}_{lk}(T)&=-\int\limits_{0}^{T}\dd t\braket{\psi_{k}|\hat{H}_{1}|\psi_{l}}e^{\ii(E_{k}-E_{l})t}+\hat{L}_{lk}(0).
\end{align}

Herein $\ket{\tilde{\psi}_{T,f}}$ refers to the dynamically changing dressed state corresponding to the state $\ket{\psi_{f}}$ and evaluated at time $T$. $\ket{\tilde{\psi}_{i}(T)}$ refers to the dressed state of $\ket{\psi_{i}}$, evaluated at time $t=0$ and time evolved under the particular model to time $T$. 

\subsection{Inner product}
Observe that
\begin{align}
\hat{L}_{lk}^{\dagger}=\hat{L}_{kl}.
\end{align}
Now
\begin{align}
\braket{\phi_{f},\tilde{\psi}_{T,f}|\tilde{\psi}_{i}(T),\phi_{i}}=\ii q\bra{\phi_{f}}\left(\hat{K}_{if}(T)e^{-\ii E_{f}T}-\hat{L}_{if}(T)e^{-\ii E_{i}T}\right)\ket{\phi_{i}},
\end{align}
where $\ket{\phi_{i,f}}$ correspond to the initial and final states of the EM field. Since there is no $O(1)$ term there is no need to keep track of the $O(q^{2})$ terms. Expanding:
\begin{align}
\braket{\tilde{\psi}_{T,f},\phi_{f}|\tilde{\psi}_{i}(T),\phi_{i}}&=\ii q e^{-\ii E_{f}T}\bra{\phi_{f}}\left(-\int\limits_{0}^{T}\dd t\braket{\psi_{f}|\hat{H}_{1}|\psi_{i}}e^{\ii(E_{f}-E_{i})t}+\hat{L}_{if}(0)-\hat{L}_{if}(T)e^{\ii(E_{f}-E_{i})T}\right)\ket{\phi_{i}},
\end{align}
let $\Omega=E_{f}-E_{i}$,
\begin{align}
\braket{\tilde{\psi}_{T,f},\phi_{f}|\tilde{\psi}_{i}(T),\phi_{i}}&=\ii q e^{-\ii E_{f}T}\bra{\phi_{f}}\left(-\int\limits_{0}^{T}\dd t\braket{\psi_{f}|\hat{H}_{1}|\psi_{i}}e^{\ii\Omega t}+\hat{L}_{if}(0)-\hat{L}_{if}(T)e^{\ii\Omega T}\right)\ket{\phi_{i}}.
\end{align}



\subsection{Transition probability}

Consider the inner product
\begin{align}
\braket{\tilde{\xi}_{f},\phi_{f}|\tilde{\xi}_{i},\phi_{i}}&=\braket{\phi_{f}|\hat{O}|\phi_{i}},
\end{align}
Then the probability will be given by
\begin{align}
\abs{\braket{\tilde{\xi}_{f},\phi_{f}|\tilde{\xi}_{i},\phi_{i}}}^{2}&=\braket{\phi_{i}|\hat{O}^{\dagger}|\phi_{f}}\braket{\phi_{f}|\hat{O}|\phi_{i}}.
\end{align}
In particular since our attention is on the transition amplitudes of the atom itself we need to trace out the final state of the field $\ket{\phi_{f}}$, leading to 
\begin{align}
\sum_{\phi_{f}}\abs{\braket{\tilde{\xi}_{f},\phi_{f}|\tilde{\xi}_{i},\phi_{i}}}^{2}&=
\bra{\phi_{i}}\hat{O}^{\dagger}\sum_{\phi_{f}}\ket{\phi_{f}}\bra{\phi_{f}}\hat{O}\ket{\phi_{i}},\\
\bra{\phi_{i}}\hat{O}^{\dagger}\mathbb{I}\hat{O}\ket{\phi_{i}}
&= \braket{\phi_{i}|\hat{O}^{\dagger}\hat{O}|\phi_{i}},
\end{align}
i.e. independent of $\phi_{f}$. Therefore in the derivation that follows $\phi_{f}$ has already been traced out and as such the notation will only make reference to $\phi_{i}$.

\begin{align}
\begin{split}
\bra{\phi_{i}}\abs{\braket{\tilde{\psi}_{T,f}|\tilde{\psi}_{i}(T)}}^{2}\ket{\phi_{i}}&=q^{2}\bra{\phi_{i}}\Bigg(
\iint\limits_{0}^{T}\dd t\, \dd t'\,\braket{\psi_{f}|\hat{H}_{1}(t)|\psi_{i}}\braket{\psi_{i}|\hat{H}_{1}(t')|\psi_{f}}e^{\ii\Omega(t-t')}\\
-&\int\limits_{0}^{T}\braket{\psi_{f}|\hat{H}_{1}|\psi_{i}}e^{\ii\Omega t}]\hat{L}_{fi}(0)
-\int\limits_{0}^{T}\braket{\psi_{i}|\hat{H}_{1}|\psi_{f}}e^{-\ii\Omega t}\hat{L}_{if}(0)\\
+&\int\limits_{0}^{T}\braket{\psi_{f}|\hat{H}_{1}|\psi_{i}}e^{\ii\Omega t}\hat{L}_{fi}(T)e^{-\ii\Omega T}
+\int\limits_{0}^{T}\braket{\psi_{i}|\hat{H}_{1}|\psi_{f}}e^{-\ii\Omega t}\hat{L}_{if}(T)e^{\ii\Omega T}\\
+&\hat{L}_{if}(0)\hat{L}_{fi}(0)+\hat{L}_{if}(T)\hat{L}_{fi}(T)-\hat{L}_{if}(0)\hat{L}_{fi}(T)e^{-\ii\Omega T}-\hat{L}_{fi}(0)\hat{L}_{if}(T)e^{\ii\Omega T}
\Bigg)\ket{\phi_{i}}
\end{split}
\end{align}

\subsubsection{Minimal model}
\begin{align}
\begin{split}
&\bra{\phi_{i}}\abs{\braket{\tilde{\psi}_{T,f}|\tilde{\psi}_{i}(T)}}^{2}\ket{\phi_{i}}\\
&=\frac{q^{2}}{\mu_{e}^{2}}\bra{\phi_{i}}\iint\limits_{-\infty}^{\infty}\dd^{3} \bm{x}\dd^{3} \bm{x}'\Bigg(
\iint\limits_{0}^{T}\dd t\, \dd t'\,
\psi_{f}^{*}(x)\left(-\ii\partial_{a}\right)\psi_{i}(x)\psi_{i}^{*}(x')\left(-\ii\partial'_{b}\right)\psi_{f}(x')\hat{A}^{a}(x,t)\hat{A}^{b}(x',t')
e^{\ii\Omega(t-t')}\\
-&\int\limits_{0}^{T}\dd t
\psi_{f}^{*}(x)\left(-\ii\partial_{a}\right)\psi_{i}(x)\hat{A}^{a}(x,t)e^{\ii\Omega t}
\psi_{i}^{*}(x')\left(-\ii\partial_{b}\right)\psi_{f}(x')\hat{A}^{b}(x',0)\frac{1}{-\ii \Omega}\\
-&\int\limits_{0}^{T}\dd t
\psi_{i}^{*}(x)\left(-\ii\partial_{a}\right)\psi_{f}(x)\hat{A}^{a}(x,t)e^{-\ii\Omega t}
\psi_{f}^{*}(x')\left(-\ii\partial_{b}\right)\psi_{i}(x')\hat{A}^{b}(x',0)\frac{1}{\ii \Omega}\\
+&\int\limits_{0}^{T}\dd t
\psi_{f}^{*}(x)\left(-\ii\partial_{a}\right)\psi_{i}(x)\hat{A}^{a}(x,t)e^{\ii\Omega t}
\psi_{i}^{*}(x')\left(-\ii\partial_{b}\right)\psi_{f}(x')\hat{A}^{b}(x',T)\frac{e^{-\ii\Omega T}}{-\ii \Omega}\\
+&\int\limits_{0}^{T}\dd t
\psi_{i}^{*}(x)\left(-\ii\partial_{a}\right)\psi_{f}(x)\hat{A}^{a}(x,t)e^{-\ii\Omega t}
\psi_{f}^{*}(x')\left(-\ii\partial_{b}\right)\psi_{i}(x')\hat{A}^{b}(x',T)\frac{e^{\ii\Omega T}}{\ii \Omega}\\
+&\psi_{f}^{*}(x)\left(-\ii\partial_{a}\right)\psi_{i}(x)\psi_{i}^{*}(x')\left(-\ii\partial_{b}\right)\psi_{f}(x')\bigg[\hat{A}^{a}(x,0)\hat{A}^{b}(x',0)\frac{1}{\Omega^{2}}
+\hat{A}^{a}(x,T)\hat{A}^{b}(x',T)\frac{1}{\Omega^{2}}\\
-&\hat{A}^{a}(x,0)\hat{A}^{b}(x',T)\frac{e^{-\ii\Omega T}}{\Omega^{2}}
-\hat{A}^{a}(x,T)\hat{A}^{b}(x',0)\frac{e^{\ii\Omega T}}{\Omega^{2}}\bigg]
\Bigg)\ket{\phi_{i}},
\end{split}
\end{align}
where the derivative operators only act on the wavefunction immediately to its right.

\begin{align}
\begin{split}
\abs{\braket{\tilde{\psi}_{T,f}|\tilde{\psi}_{i}(T)}}^{2}&=\frac{q^{2}}{\mu_{e}^{2}}\iint\limits_{-\infty}^{\infty}\dd^{3} \bm{x}\dd^{3} \bm{x}'\Bigg(
\iint\limits_{0}^{T}\dd t\, \dd t'\,
\psi_{f}^{*}(x)\left(-\ii\partial_{a}\right)\psi_{i}(x)\psi_{i}^{*}(x')\left(-\ii\partial'_{b}\right)\psi_{f}(x')e^{\ii\Omega(t-t')}\bigg<\hat{A}^{a}(x,t)\hat{A}^{b}(x',t')\bigg>_{\phi_{i}}
\\
-&\int\limits_{0}^{T}\dd t
\psi_{f}^{*}(x)\left(-\ii\partial_{a}\right)\psi_{i}(x)e^{\ii\Omega t}
\psi_{i}^{*}(x')\left(-\ii\partial_{b}\right)\psi_{f}(x')\frac{1}{-\ii \Omega}\bigg<\hat{A}^{a}(x,t)\hat{A}^{b}(x',0)\bigg>_{\phi_{i}}\\
-&\int\limits_{0}^{T}\dd t
\psi_{i}^{*}(x)\left(-\ii\partial_{a}\right)\psi_{f}(x)e^{-\ii\Omega t}
\psi_{f}^{*}(x')\left(-\ii\partial_{b}\right)\psi_{i}(x')\frac{1}{\ii \Omega}\bigg<\hat{A}^{b}(x',0)\hat{A}^{a}(x,t)\bigg>_{\phi_{i}}\\
+&\int\limits_{0}^{T}\dd t
\psi_{f}^{*}(x)\left(-\ii\partial_{a}\right)\psi_{i}(x)e^{\ii\Omega t}
\psi_{i}^{*}(x')\left(-\ii\partial_{b}\right)\psi_{f}(x')\frac{e^{-\ii\Omega T}}{-\ii \Omega}\bigg<\hat{A}^{a}(x,t)\hat{A}^{b}(x',T)\bigg>_{\phi_{i}}\\
+&\int\limits_{0}^{T}\dd t
\psi_{i}^{*}(x)\left(-\ii\partial_{a}\right)\psi_{f}(x)e^{-\ii\Omega t}
\psi_{f}^{*}(x')\left(-\ii\partial_{b}\right)\psi_{i}(x')\frac{e^{\ii\Omega T}}{\ii \Omega}\bigg<\hat{A}^{b}(x',T)\hat{A}^{a}(x,t)\bigg>_{\phi_{i}}\\
+&\psi_{f}^{*}(x)\left(-\ii\partial_{a}\right)\psi_{i}(x)\psi_{i}^{*}(x')\left(-\ii\partial_{b}\right)\psi_{f}(x')\frac{1}{\Omega^{2}}\bigg<\hat{A}^{a}(x,0)\hat{A}^{b}(x',0)
+\hat{A}^{a}(x,T)\hat{A}^{b}(x',T)\\
-&\hat{A}^{a}(x,0)\hat{A}^{b}(x',T)e^{-\ii\Omega T}
-\hat{A}^{a}(x,T)\hat{A}^{b}(x',0)e^{\ii\Omega T}\bigg>_{\phi_{i}}
\Bigg),
\end{split}\label{eqc26}
\end{align}
where the final step was to ensure that the operators $\hat{\bm{A}}$ were all in the correct order (i.e. $\braket{\psi_{f}|\psi_{i}}$, f before i). Note that this order is reversible however it must be consistent.

\subsubsection{Dipole model}
\begin{align}
\begin{split}
\abs{\braket{\tilde{\psi}_{T,f}|\tilde{\psi}_{i}(T)}}^{2}&=q^{2}\iint\limits_{-\infty}^{\infty}\dd^{3} \bm{x}\dd^{3} \bm{x}'
\iint\limits_{0}^{T}\dd t\, \dd t'\,
\psi_{f}^{*}(x)\left(r_{a}\right)\psi_{i}(x)\psi_{i}^{*}(x')\left(r'_{b}\right)\psi_{f}(x')e^{\ii\Omega(t-t')}\bigg<\hat{E}^{a}(x,t)\hat{E}^{b}(x',t')\bigg>_{\phi_{i}}.
\end{split}\label{eqc27}
\end{align}

Note that both \eqref{eqc26} and \eqref{eqc27} have been written down in this format for those who prefer to work with Wightman functions.

\section{Numerical setup}\label{seca3}

In order to evaluate the transition probabilities shown in Appendix \ref{seca2} certain simplifications are made prior to resorting to numerical integration. These simplifications have been made with the aid of computer algebra software.

\subsubsection{Minimal model}
In order to optimise computational resources start with (operator equations over EM Hilbert space)
\begin{align}
\braket{\tilde{\psi}_{T,f}|\tilde{\psi}_{i}(T)}&=\ii q e^{-\ii E_{f}T}\left(-\int\limits_{0}^{T}\dd t\braket{\psi_{f}|\hat{H}_{1}|\psi_{i}}e^{\ii\Omega t}+\hat{L}_{if}(0)-\hat{L}_{if}(T)e^{\ii\Omega t}\right),
\end{align}
and recall $\hat{H}_{1}=-\dfrac{\hat{\bm{A}}\cdot\hat{\bm{p}}}{\mu_{e}}$. Then
\begin{align}
\begin{split}
\braket{\tilde{\psi}_{T,f}|\tilde{\psi}_{i}(T)}&=
\ii\frac{q}{\mu_{e}}e^{-\ii E_{f}T}\bra{\psi_{f}}\int\frac{d^{3}k}{(2\pi)^{3/2}\sqrt{2\omega}}\sum_{\lambda=1}^{2}\bm{\epsilon}_{\lambda}(k)\left(\hat{a}_{\lambda}^{\vphantom{\dagger}}\left\{\int\limits_{0}^{T} e^{-\ii\omega t}e^{\ii\Omega t}dt+\frac{1}{\ii\Omega}-\frac{e^{-\ii\omega T}e^{\ii\Omega T}}{\ii\Omega}\right\}e^{\ii\bm{k}\cdot\bm{x}}\hat{\bm{p}}\right.\\
+&\left.
\hat{a}_{\lambda}^{\dagger}\left\{\int\limits_{0}^{T} e^{\ii\omega t}e^{\ii\Omega t}dt+\frac{1}{\ii\Omega}-\frac{e^{\ii\omega T}e^{\ii\Omega T}}{\ii\Omega}\right\}e^{-\ii\bm{k}\cdot\bm{x}}\hat{\bm{p}}\right)\ket{\psi_{i}}
\end{split}\notag\\
\begin{split}
&=\frac{q}{\mu_{e}}e^{-\ii E_{f}T}\int\frac{d^{3}k}{(2\pi)^{3/2}\sqrt{2\omega}}\sum_{\lambda=1}^{2}\bm{\epsilon}_{\lambda}(k)\left(\hat{a}_{\lambda}^{\vphantom{\dagger}}e^{\ii(\Omega-\omega)\frac{T}{2}}\left\{T\frac{2\sin\left((\Omega-\omega)\frac{T}{2}\right)}{(\Omega-\omega)T}-\frac{2\sin\left((\Omega-\omega)\frac{T}{2}\right)}{\Omega}\right\}\right.\\ \times&\braket{\psi_{f}|e^{\ii\bm{k}\cdot\bm{x}}\nabla|\psi_{i}}
+\left.
\hat{a}_{\lambda}^{\dagger}e^{\ii(\Omega+\omega)\frac{T}{2}}\left\{T\frac{2\sin\left((\Omega+\omega)\frac{T}{2}\right)}{(\Omega+\omega)T}-\frac{2\sin\left((\Omega+\omega)\frac{T}{2}\right)}{\Omega}\right\}\braket{\psi_{f}|e^{-\ii\bm{k}\cdot\bm{x}}\nabla|\psi_{i}}\right)
\end{split}\notag\\
\begin{split}
&=\frac{q}{\mu_{e}}e^{-\ii E_{f}T}\int\frac{d^{3}k}{(2\pi)^{3/2}}\sqrt{\frac{\omega}{2}}\sum_{\lambda=1}^{2}\bm{\epsilon}_{\lambda}(k)\left(\hat{a}_{\lambda}^{\vphantom{\dagger}}e^{\ii(\Omega-\omega)\frac{T}{2}}\left\{T\frac{2\sin\left((\Omega-\omega)\frac{T}{2}\right)}{(\Omega-\omega)T}\right\}\braket{\psi_{f}|\frac{e^{\ii\bm{k}\cdot\bm{x}}\nabla}{\Omega}|\psi_{i}}\right.\\
-&\left.
\hat{a}_{\lambda}^{\dagger}e^{\ii(\Omega+\omega)\frac{T}{2}}\left\{T\frac{2\sin\left((\Omega+\omega)\frac{T}{2}\right)}{(\Omega+\omega)T}\right\}\braket{\psi_{f}|\frac{e^{-\ii\bm{k}\cdot\bm{x}}\nabla}{\Omega}|\psi_{i}}\right)
\end{split}\notag\\
\begin{split}
&=T\frac{q}{\mu_{e}}e^{-\ii E_{f}T}\int\frac{d^{3}k}{(2\pi)^{3/2}}\sqrt{\frac{\omega}{2}}\sum_{\lambda=1}^{2}\bm{\epsilon}_{\lambda}(k)\left(\hat{a}_{\lambda}^{\vphantom{\dagger}}e^{\ii(\Omega-\omega)\frac{T}{2}}\text{sinc}\left((\Omega-\omega)\frac{T}{2}\right)\braket{\psi_{f}|\frac{e^{\ii\bm{k}\cdot\bm{x}}\nabla}{\Omega}|\psi_{i}}\right.\\
-&\left.
\hat{a}_{\lambda}^{\dagger}e^{\ii(\Omega+\omega)\frac{T}{2}}\text{sinc}\left((\Omega+\omega)\frac{T}{2}\right)\braket{\psi_{f}|\frac{e^{-\ii\bm{k}\cdot\bm{x}}\nabla}{\Omega}|\psi_{i}}\right). 
\end{split}
\end{align}

\subsubsection{Dipole model}

\begin{align}
\begin{split}
\braket{\tilde{\psi}_{T,f}|\tilde{\psi}_{i}(T)}&=
-Tqe^{-\ii E_{f}T}\int\frac{d^{3}k}{(2\pi)^{3/2}}\sqrt{\frac{\omega}{2}}\sum_{\lambda=1}^{2}\bm{\epsilon}_{\lambda}(k)\left(\hat{a}_{\lambda}^{\vphantom{\dagger}}e^{\ii(\Omega-\omega)\frac{T}{2}}\text{sinc}\left((\Omega-\omega)\frac{T}{2}\right)\braket{\psi_{f}|e^{\ii\bm{k}\cdot\bm{x}}\hat{\bm{r}}|\psi_{i}}\right.\\
-&\left.\hat{a}_{\lambda}^{\dagger}
e^{\ii(\Omega+\omega)\frac{T}{2}}\text{sinc}\left((\Omega+\omega)\frac{T}{2}\right)\braket{\psi_{f}|e^{-\ii\bm{k}\cdot\bm{x}}\hat{\bm{r}}|\psi_{i}}\right). 
\end{split}
\end{align}

In both cases expressions of the form $\begin{aligned}[t]  \braket{\psi_{f}|e^{-\ii\bm{k}\cdot\bm{x}}\nabla|\psi_{i}}, \braket{\psi_{f}|e^{-\ii\bm{k}\cdot\bm{x}}\hat{\bm{r}}|\psi_{i}}\end{aligned}$ need to be evaluated. Firstly we note that
\begin{align}
\braket{\psi_{f}|\hat{O}|\psi_{i}}&=\int\dd^{3}\bm{x}\psi_{f}^{*}(\bm{x})O(\bm{x},\partial_{\bm{x}})\psi_{i}(\bm{x})\notag\\
&=\int\dd^{3}\bm{x}R_{n_{f},l_{f}}^{*}(r)Y^{*}_{l_{f},m_{f}}(\hat{\bm{x}})O(\bm{x},\partial_{\bm{x}})
R_{n_{i},m_{i}}(r)Y_{l_{i},m_{i}}(\hat{\bm{x}}).
\end{align}
This final expression above can be simplified by exploiting Clebsch-Gordon coefficients in reducing the product of spherical harmonics into a sum of spherical harmonics. Importantly this will be a finite sum. Secondly, note that
\begin{align}
e^{\ii \bm{x}\cdot\bm{k}}=\sum_{l=0}^{\infty}\sum_{m=-l}^{l}4\pi\ii^{l}j_{l}(\abs{\bm{x}}\abs{\bm{k}})Y_{lm}(\hat{\bm{x}})Y^{*}_{lm}(\hat{\bm{k}})=\sum_{l=0}^{\infty}\sum_{m=-l}^{l}4\pi\ii^{l}j_{l}(\abs{\bm{x}}\abs{\bm{k}})Y^{*}_{lm}(\hat{\bm{x}})Y_{lm}(\hat{\bm{k}}).
\end{align}
Using this identity our desired terms reduce to $\psi_{f}^{*}\nabla\psi_{i}\rightarrow\sum_{(\lambda,\mu)\in S}\bm{u}_{\lambda\mu}Y_{\lambda\mu}$, therefore
\begin{align}
\braket{\psi_{f}|e^{-\ii\bm{k}\cdot\bm{x}}\nabla|\psi_{i}}&=\int\dd^{3}\bm{x}\sum_{(\lambda,\mu)\in S}\bm{u}_{\lambda \mu}Y_{\lambda\mu}(\hat{\bm{x}})\sum_{l=0}^{\infty}\sum_{m=-l}^{l}4\pi\ii^{l}j_{l}(\abs{\bm{x}}\abs{\bm{k}})Y^{*}_{lm}(\hat{\bm{x}})Y_{lm}(\hat{\bm{k}})\notag\\
&=\sum_{(\lambda,\mu)\in S}\sum_{l=0}^{\infty}\sum_{m=-l}^{l}4\pi\ii^{l}\bm{u}_{\lambda \mu}j_{l}(\abs{\bm{x}}\abs{\bm{k}})Y_{lm}(\hat{\bm{k}})\delta_{\lambda l}\delta_{\mu m}\notag\\
&=\sum_{(\lambda,\mu)\in S}4\pi\ii^{\lambda}\bm{u}_{\lambda \mu}j_{\lambda}(\abs{\bm{x}}\abs{\bm{k}})Y_{\lambda\mu}(\hat{\bm{k}}),
\end{align}
where $S$ is a finite set. Note that whilst this derivation may seem abstract and ineffective, these steps can be followed with computer algebra packages. The above can be interpreted as an illustrated pseudo-code.

The remainder of the calculation is straightforward, requiring eliminating the EM field creation and annihilation operators via inner products and a series of $\bm{k}$ space integrals, of which the angular parts can be evaluated analytically and the radial part $\omega$ must be evaluated numerically except for the special cases of rotating wave approximation or for very long times $T\gg a_{0}/Z$.

\end{widetext}
\bibliography{bibliography}

\end{document}